\documentclass{article}

\usepackage{arxiv}

\usepackage[utf8]{inputenc} 
\usepackage[T1]{fontenc}    
\usepackage{hyperref}       
\usepackage{url}            
\usepackage{booktabs}       
\usepackage{amsfonts}       
\usepackage{nicefrac}       
\usepackage{microtype}      
\usepackage{lipsum}
\usepackage{gensymb}
\usepackage{longtable}
\usepackage{wrapfig}
\usepackage{soul}
\usepackage{booktabs}
\usepackage{graphicx}
\usepackage{float}
\restylefloat{table}
\usepackage{amsmath}
\usepackage{subcaption}
\usepackage{bm}
\usepackage{appendix}
\usepackage{multirow}
\usepackage[table,xcdraw]{xcolor}
\usepackage{longtable}
\usepackage{listings}
\usepackage{subcaption}
\usepackage{natbib}

\title{Can Machine Learning Identify Governing Laws for Dynamics in Complex Engineered Systems ? : A Study in Chemical Engineering}

\author{
  Renganathan~Subramanian\thanks{Renganathan is a Senior UG Student at IIT-M. This research was done as Visiting Student at Purdue University.} \\
  Department of Chemical Engineering\\
  Indian Institute of Technology, Madras\\
  Madras, India \\
  \texttt{ch16b058@smail.iitm.ac.in} \\
   \And 
Shweta~Singh\thanks{Shweta Singh is an Assistant Professor at Purdue University, Corresponding Author} \\
  Department of Agricultural \& Biological Engineering\\
  Division of Environmental \& Ecological Engineering\\
  Purdue University \\
  West Lafayette, IN, USA\\
  \texttt{singh294@purdue.edu} \\
}

\begin{document}
\maketitle

\begin{abstract}
Machine learning recently has been used to identify the governing equations for dynamics in physical systems. The promising results from applications on systems such as fluid dynamics and chemical kinetics inspire further investigation of these methods on complex engineered systems. While at micro-scales the governing laws such as heat transfer, diffusion, pressure variation are well known and have been studied for decades, it is also known that the laws for complex systems are not well established. Dynamics of these systems play a crucial role in design and operations. Hence, it would be advantageous to learn more about the mechanisms that may be driving the complex dynamics of systems where an overall governing law is unknown. In this work, our research question was aimed at addressing this open question about applicability and usefulness of novel machine learning approach in identifying the governing dynamical equations for engineered systems. We focused on distillation column which is an ubiquitous unit operation in chemical engineering and demonstrates complex dynamics i.e. it's dynamics is a combination of heuristics and fundamental physical laws. We tested the method of Sparse Identification of Non-Linear Dynamics (SINDy) because of it's ability to produce white-box models with terms that can be used for physical interpretation of dynamics. Dynamics of the system was externally forced using perturbations to input stream and time series data was generated from simulation of distillation column using ASPEN Dynamics software. One promising result was reduction of number of equations for dynamic simulation from 1000s in ASPEN to only 13 - one for each state variable. Prediction accuracy was high on the test data from system within the perturbation range, however outside perturbation range equations did not perform well. In terms of physical law extraction, some terms were interpretable as related to Fick's law of diffusion (with concentration terms) and Henry's law (with ratio of concentration and pressure terms). Equations were complex, some terms were interpretable but we did not achieve a conclusive answer on physical governing laws. We conclude that more research is needed on combining engineering systems with machine learning approach to improve understanding of unknown dynamics. 
\end{abstract}

\keywords{Machine Learning \and Chemical Engineering \and ASPEN Dynamics \and Distillation Column \and SINDy \and Dynamic Equations}

\section{Introduction}
Engineering has relied on identification of system dynamics from first principle methods for decades in order to understand the underlying the mechanisms driving dynamics.These first principle equations form the fundamentals that are used to design and operate systems for desired outputs such as heat transfer, operation of process plants, fluid flow operations etc. The equations are also augmented with observation driven empirical relationships which are not fundamentally a law but form basic governing rules. Combined both mechanistic and empirical rules form core of engineering in design and operations.

However, in several cases the first principle based equations may not be available for the system or might be extremely complicated for quick computation and analysis. In these scenarios, it becomes necessary to develop an understanding of the system using data-driven methods. With the advancement in tools to generate, store, transport and analyze high quality and high quantity data, it has become inevitable to rely on data-driven methods to extract governing equations and patterns for a system. However the use of data-driven methods to understand dynamical systems has been limited. In chemical engineering systems a single unit can exhibit complex dynamics. This motivates the need to explore the potential of data based methods to extract dynamic governing equations in such systems.

In this work, we try to identify system dynamics of a distillation column. Distillation columns are one of the well studied and established units in Chemical Engineering. We first build a dynamic process flow simulation of this distillation column to generate time series data. We then apply data driven system identification on this system and try to answer the question of whether data based machine learning methods can replace first principles.

The paper is organized as follows. In Section \ref{sec:sparse} we explain the method of Sparse Regression of Dynamical Systems. This method has been shown \cite{Brunton3932, hoffmann2019} to possess the ability to extract sparse governing equations for dynamic systems and can balance model complexity with model accuracy. In Section \ref{subsec:column}, we explain the process flow simulation development on Aspen Plus\textsuperscript{\textregistered} for the distillation column. The rest of Section \ref{sec:method} deals with creating a dynamic simulation, data generation, model training, selection and testing. In Section \ref{sec:results} we show the results of the research and interpret them with the goal to establish the extent to which machine learning can substitute for first principles. We finally discuss the key takeaways and the prospects for future research in Section \ref{sec:discuss}

\section{Approach : Physical System Identification Using Sparse Identification of Non-Linear Dynamics (SINDy)}

\subsection{Method of Sparse Identification of Non-Linear Dynamics}
\label{sec:sparse}

Sparse regression is a machine learning methodology which works under the assumption that the governing equations of most dynamical systems contain very few terms. These equation can be considered sparse in the function space and the system is expected to evolve on a low dimensional manifold. Sparse Identification tries to discover these equations from noisy time series data.\\
We consider systems whose governing equations are non-linear ODEs of the form,\\
$$
\frac{d\bm{x(t)}}{dt} = f(\bm{x(t)}, u(t))
$$
where $\bm{x(t)} \in \mathbb{R}^n$ denotes the state of the system at time $t$, $u(t) \in \mathbb{R}$ is the input function value at time $t$ and  $f\left(x(t)\right)$ is a linear combination of non-linear functions of $\bm{x(t)}$ and $u(t)$. Mathematically,
\begin{align*}
\bm{\dot{x}(t)} &=
\displaystyle\Sigma_{i=1}^k\xi_i\theta_i\left(\bm{x(t)}, u(t)\right)\\
\bm{x(t)} &= \begin{bmatrix}
x_1(t) & x_2(t) & \hdots x_n(t)
\end{bmatrix}
\end{align*}
Where $x_1, \hdots x_n$ are the states of the system, $\theta$s are non-linear functions called the candidate terms of $f(x)$, and $\xi$s are the coefficients of the terms. We expect most of $\xi$s to be $0$ making $f(x, u)$ sparse in the number of terms. The goal of the algorithm is to identify the very few terms which make up $f(x, u)$ from a very large set of candidates. Instead of a combinatorial search for these terms by brute force, the algorithm includes a penalty for model complexity. This forces the selected function to be sparse. By forcing sparse functions the algorithm also ensures that the model obtained does not overfit the data.\\
The data required for the algorithm is a time series of states arranged in a matrix $\bm{X(t)} \in \mathbb{R}^{m\times n}$ of the form
\begin{align*}
    \bm{X(t)} = \begin{bmatrix}
        x_1(t) & x_2(t) & \hdots & x_n(t)\\
        x_1(t-1) & x_2(t-1) & \hdots & x_n(t-1)\\
        \vdots &  & &\vdots\\
        x_1(t-m+1) & x_2(t-m+1) & \hdots & x_n(t-m+1)
    \end{bmatrix}
\end{align*}
The derivative of $\bm{X(t)}, \bm{\dot{X}(t)} \in \mathbb{R}^{m\times n}$ is matrix of the form
\begin{align*}
    \bm{\dot{X}(t)} = \begin{bmatrix}
        \dot{x_1}(t) & \dot{x_2}(t) & \hdots & \dot{x_n}(t)\\
        \dot{x_1}(t-1) & \dot{x_2}(t-1) & \hdots & \dot{x_n}(t-1)\\
        \vdots &  & &\vdots\\
        \dot{x_1}(t-m+1) & \dot{x_2}(t-m+1) & \hdots & \dot{x_n}(t-m+1)
    \end{bmatrix}
\end{align*}
obtained by numerically differentiating $X(t)$. And,
$$
\bm{u(t)} = 
\begin{bmatrix} u(t), u(t-1), \hdots u(t-m+1) \end{bmatrix}^T
$$
The governing equation becomes
$$
    \bm{\dot{X}(t)} = \Theta\left(\bm{X(t), u(t)}\right)\Xi
$$
    With $\Theta\left(\bm{X(t)}\right) \in \mathbb{R}^{m\times k}$ given by 
\begin{math}
    \begin{bmatrix}
    \theta_1\left(\bm{X(t)}\right) & \theta_2\left(\bm{X(t)}\right) \hdots
    \end{bmatrix}
\end{math}\\
And $\Xi \in \mathbb{R}^{k\times n}$ given by
$ \begin{bmatrix}
        \xi_1 & \xi_2 & \hdots \xi_n
    \end{bmatrix} $\\~\\
We try to identify the sparse matrix $\Xi$ by solving the least squares optimization problem. However, this includes an optimization for every column of $\bm{\dot{X}(t)}$. So in this case, we have to solve n optimization problems, one for each of the n states of the system.\\
The algorithm forces sparsity by adding a regularization term to the objective function. The ideal regularization to force sparsity would be minimizing the $L_0$ norm of the coefficients (number of non zero terms in the vector). But this an NP-hard problem \cite{natarjan1995sparse}. However, it has been shown \cite{Donoho2197} that mininmizing the $L_1$ norm is a convex optimization and also produces solutions which are sparse. This is referred to as the lease absolute shrinkage and selection operator (LASSO). The LASSO optimization problem is
$$
    \xi_i^* = \underset{\xi_i}{\text{argmin }}\left||\bm{\dot{x_i}}-\Theta\left(\bm{X(t)}\right)\xi_i\right||_2+\alpha||\xi_i||_1 \quad \quad i = 1,2 \hdots n
$$
$\alpha$ is the regularization parameter which has to be tuned inorder to achieve a trade-off between accuracy and sparsity. This optimization problem can be solved by the standard convex optimization algorithms. We have used coordinate descent algorithm which is available as a prebuilt function in the \textit{scikit} Python library.\\
The capability of the algorithm to capture the dynamics of the system depends mainly on the candidate functions provided. Some prior knowledge of the process might help identify these candidate functions. This is a place where domain knowledge becomes important. The method also depends on the quality of the data. Therefore, we need to filter the derivatives and/or variables as we are using numerical differentiation to obtain the derivatives.
\subsection{Physical System for Identification : A Distillation Column}
\label{subsec:column}
In order to study the application of machine learning approach to identify the governing equations for dynamics in engineered systems we selected the unit operation of distillation column. Distillation columns are one of the most ubiquitous unit operations in process industries ranging from petrochemicals, biomass to now the next generation biorefineries. While the column looks simple from the operations perspective (after multiple decades of theory and design development), the dynamics of this system is complex. The dynamics of the system is dependent on multiple physical laws such as heat transfer, diffusion principles, mass flow dynamics, hueristics that relate the pressure to chemical properties, temperature and pressure relationships etc. A standard software used to model the operation of such an unit can include upto 1000s of equation. While control principles using linearized models are already  being used to deverlop control systems for these units, an overall simple law that govern the dynamics of these systems is not known. Our goal of using machine learning based approach was to test the applicability of simplified data driven approach to identify the governing laws as data can be more easily generated. First principles approach to identify complex dynamics of these systems will certainly be a much difficult task. Next we describe the system selected, generation of time series data and selected system variables that describe the state of system to apply SINDy method.

\subsubsection{Test Distillation Column} \label{distColumnStudied}
\begin{figure}[H]
\centering
  \includegraphics[width=0.5\linewidth]{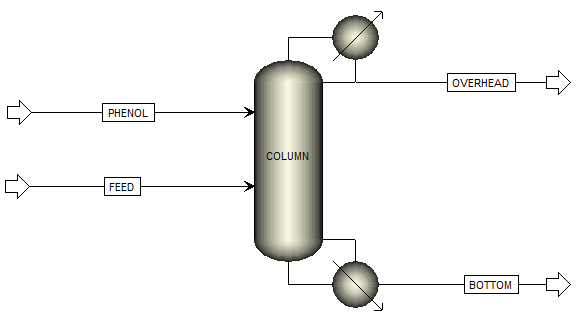}
  \caption{Process Flow Diagram - Distillation Column}
  \label{fig:column}
\end{figure}

The system considered was an extractive distillation column used to recover methylcyclohexane (MCH) from a mixture of MCH and toulene. Since MCH (Boiling Point = $101 \degree$ C) and toluene (Boiling Point = $110.6 \degree$ C) have very close boiling points they cannot be separated by a conventional distillation column. Therefore, we use phenol (Boiling Point = $181.7 \degree$ C) which has a higher affinity towards toluene to alter the relative volatility and promote separation. An equimolar mixture of MCH and toluene forming the feed stream and a pure phenol stream are fed to the distillation column . MCH is extracted as the overhead product while toluene and phenol leave as the bottoms products. The process flow diagram for the column is given in Fig.\ref{fig:column}. The column was modelled as a RadFrac unit. The specifications of the distillation column used are listed in Table.\ref{tab:column_specs} and the feed conditions are given in Table.\ref{tab:feed_specs}
The column is able to recover $97.3\%$ of the MCH originally present in the feed stream. The process flow diagram is then exported to Aspen Dynamics for running dynamic simulations that can allow extracting the rules governing dynamics for this system. The first-principle based mechanistic model has 2403 variables and 1848 equations as identified by Aspen Dynamics, however structure of all these equations are not known.
\begin{longtable}[h!]{|ll|r|}
    \caption{Distillation Column Specifications}\\
    
    \hline
    \multicolumn{2}{|c|}{\textbf{Specification}} & \multicolumn{1}{l|}{Value} \\ \hline
   
    %
    %
    No. of stages &  & 22 \\ \hline
    Reflux Ratio &  & 8 \\ \hline
    Distillate Rate (lbmol/hr) &  & 200 \\ \hline
    FEED Stage &  & 14 \\ \hline
    PHENOL Stage &  & 7 \\ \hline
    Stage 1 (Condenser) Pressure (psia) &  & 16 \\ \hline
    Stage 22 (Reboiler) Pressure (psia) &  & 20.2 \\ \hline
    \multicolumn{1}{|l|}{} & Diameter (ft) & 5 \\ \cline{2-3} 
    \multicolumn{1}{|l|}{} & Spacing (ft) & 2 \\ \cline{2-3} 
    \multicolumn{1}{|l|}{} & Weir Height (ft) & 0.164 \\ \cline{2-3} 
    \multicolumn{1}{|l|}{} & Lw/D & 0.7267 \\ \cline{2-3} 
    \multicolumn{1}{|l|}{Tray Geometry} & \% Active Area & 90 \\ \cline{2-3} 
    \multicolumn{1}{|l|}{} & Overall Efficiency & 1 \\ \cline{2-3} 
    \multicolumn{1}{|l|}{} & \% Hole Area & 10 \\ \cline{2-3} 
    \multicolumn{1}{|l|}{} & Hole Diameter (ft) & 0.0833 \\ \cline{2-3} 
    \multicolumn{1}{|l|}{} & \% Downcomer Escape Area & 10 \\ \cline{2-3} 
    \multicolumn{1}{|l|}{} & Foaming Factor & 1 \\ \hline
    \multicolumn{1}{|l|}{} & Length (ft) & 6 \\ \cline{2-3} 
    \multicolumn{1}{|l|}{Reflux Drum} & Diameter (ft) & 3 \\ \cline{2-3}
    \multicolumn{1}{|l|}{} & Head Type & Horizontal \\ \hline
    \multicolumn{1}{|l|}{} & Height (ft) & 5 \\ \cline{2-3} 
    \multicolumn{1}{|l|}{Sump} & Diameter (ft) & 3 \\ \cline{2-3} 
    \multicolumn{1}{|l|}{} & Head Type & Elliptical \\ \hline
    \multicolumn{1}{|l|}{} & Type & Total \\ \cline{2-3} 
    \multicolumn{1}{|l|}{} & Heat Transfer & LMTD \\ \cline{2-3} 
    \multicolumn{1}{|l|}{Condenser} & Medium Temperature (F) & 68 \\ \cline{2-3} 
    \multicolumn{1}{|l|}{} & Temperature Approach (F) & 18 \\ \cline{2-3} 
    \multicolumn{1}{|l|}{} & Heat Capacity (Btu/lb-R) & 1.00315 \\ \hline
    \multicolumn{1}{|l|}{} & Type & Kettle \\ \cline{2-3} 
    \multicolumn{1}{|l|}{Reboiler} & Heat Transfer & Constant Duty \\ \cline{2-3}
    \multicolumn{1}{|l|}{} & Heat Duty (Btu/hr) & 31615232.6 \\ \hline
    \label{tab:column_specs}
\end{longtable}

\begin{longtable}[h!]{|ll|l|l|}
    \caption{Feed Specifications}\\
    \hline
   \multicolumn{2}{|c|}{} &     PHENOL & FEED \\ \hline
    %
    %
    Molar Flow (lbmol/hr) &  & 1200 & 400 \\ \hline
    \multicolumn{1}{|l|}{} & Phenol & 1 & 0 \\ \cline{2-4} 
    \multicolumn{1}{|l|}{Mole Fraction} & Toluene & 0 & 0.5 \\ \cline{2-4} 
    \multicolumn{1}{|l|}{} & MCH & 0 & 0.5 \\ \hline
    Temperature (F) &  & 220 & 220 \\ \hline
    Pressure (psia) &  & 20 & 20 \\ \hline
     \label{tab:feed_specs}
\end{longtable}

\subsubsection{Dynamics and Time Series Generation}
\begin{figure}[H]
\centering
  \includegraphics[width=1\linewidth]{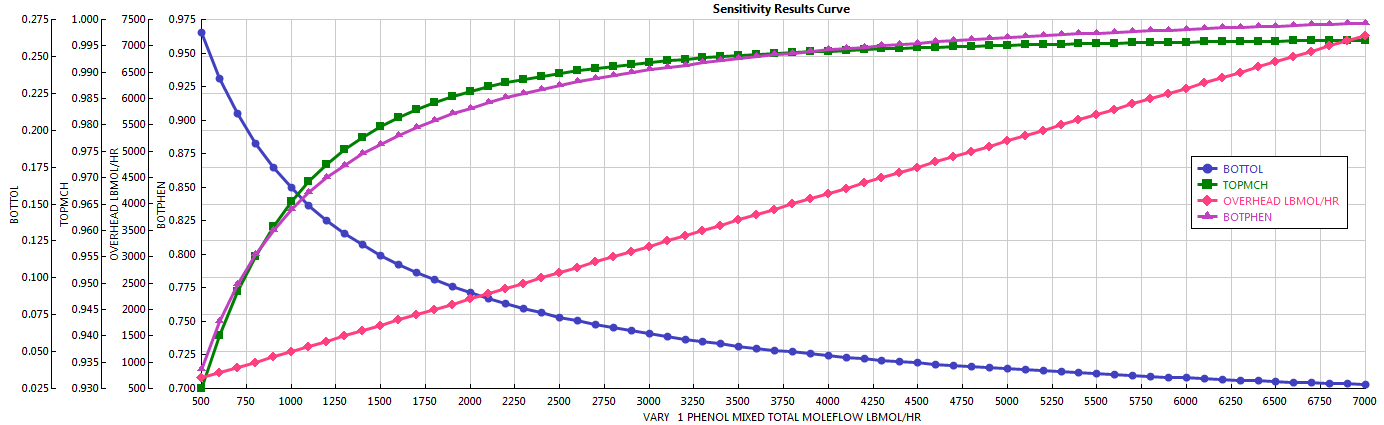}
  \caption{Sensitivity Analysis on Exit Streams for Phenol Flow Rate Variations}
  \label{fig:sensitivity}
\end{figure}

In order to capture the dynamics, system was perturbed by adding perturbations to the  phenol feed stream and the feed flow rate was kept constant. This will allow the approach to identify equations that govern the dynamics developed in the system due to changes in the extracting agent's flow rate. An initial sensitivity analysis was carried out in Aspen Plus Steady State (results shown in Fig.\ref{fig:sensitivity} to identify the valid values for phenol flow rates for which the column can operate without errors.
The perturbations were restricted to a fraction of this zone and the rest of the valid region was used for testing. \\

\begin{figure}[H]
\centering
  \includegraphics[width=1\linewidth]{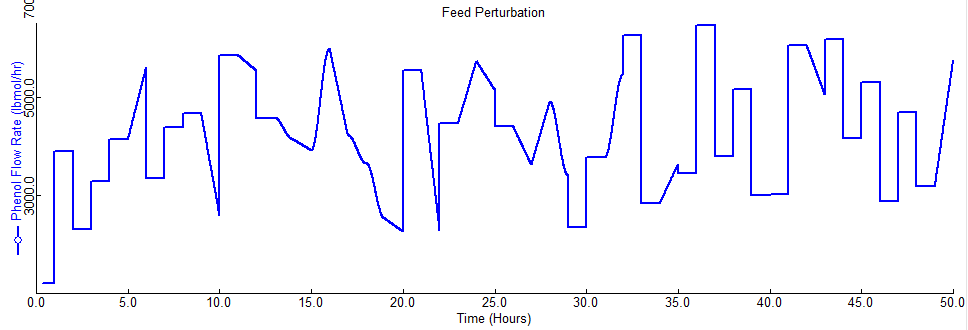}
  \caption{Perturbations - Phenol Flow Rate (lbmol/hr) vs Time (hr)}
  \label{fig:feed_perturb}
\end{figure}

\noindent{\textbf{Perturbations :}}
The phenol feed perturbation was implemented by executing a Task in Aspen Dynamics The perturbation was a random mix of step changes, linear ramps and sigmoidal ramps with a time period of 1 hour each and amplitudes between 1000 lbmol/hr to 3000 lbmol/hr generated randomly with a uniform probability distribution function. The simulation was run for 100 hours with a calculation step size of 0.01 hours. One such feed flow rate time series plot is given in Fig.\ref{fig:feed_perturb} This generates 50001 equally spaced (in time) data points. The phenol feed time series becomes $\bm{u(t)}$ which is equivalent to a forcing function that drives the dynamics of the system.

\noindent{\textbf{Operating Conditions :}}
In order to define the system, the following variables were fixed as operating condition parameters : Reflux ration, toluene feed rate, MCH feed rate, distillation column sizing, tray geometry, reboiler geometry and sizing, condenser geometry and sizing, reboiler duty and condenser heat transfer coefficients.These conditions play a crucial role in operation of selected distillation column hence fixing these parameters would allow us to identify the governing equations for mechanisms that drive the dynamics of flow streams. Further, in order to test the robustness of the equations extracted, the structure of the obtained equations were compared across different operating conditions obtained by altering these parameters. The different operating conditions tested are listed in Table.\ref{tab:op_cond}. This testing method has been further explained in Section \ref{subsec:testing}\\
\vspace{-1.0in}
\begin{table}[H]
 \caption{Different Operating Conditions Tested}
\begin{center}
    \label{tab:op_cond}
    \begin{tabular}{|l|l|l|l|l|}
    \hline
    \textbf{Parameter} & \textbf{System 1}& \textbf{System 2} & \textbf{System 3} & \textbf{System 4} \\ \hline
       \textbf{Reflux Ratio} & 6 & 10 & 8 & 8 \\ \hline
    \textbf{Toluene Feed} & 200 & 200 & 200 & 400 \\ \hline
    \textbf{MCH Feed} & 200 & 200 & 400 & 200 \\ \hline
    \end{tabular}
    \end{center}
\end{table}

\subsubsubsection{\textbf{States of the System}}
Studying the dynamics of a systems requires following the state of system by mapping the state to observable variables. In this case, for the  machine learning algorithm to capture the complete dynamics of the system, we included the set of variables which change with the perturbations and are not fixed as operating conditions. The result will be a system of ODEs that can describe the evolution of the whole system as state space dynamics for these variables. For the system under consideration, we initially chose the following variables:

\begin{table}[H]
	\begin{center}
			\begin{tabular}{|p{1.3in}|p{0.58in}|p{1.1in}|}
			\hline
		 Variables & Description & Symbol Used in ODEs \\
		 \hline
		 OVERHEAD Stream Temperature& Top T & \(TOP_{T}\)\\
		 \hline
		 OVERHEAD Stream – Phenol Flow Rate & Top Ph & ${Top}_{\text{Ph}}$\\
		 \hline
		 OVERHEAD Stream –Toluene Flow Rate & Top Tol & ${Top}_{\text{Tol}}$\\
		 \hline
		 BOTTOMS Stream - MCH Flow Rate & Bot MCH & ${Bot}_{\text{MCH}}$\\
		 \hline
		 BOTTOMS Stream – Phenol Flow Rate & Bot Ph & ${Bot}_{\text{Ph}}$ \\
		 \hline
    BOTTOMS Stream –Toluene Flow Rate &  Bot Tol & $\text{Bot}_{\text{Tol}}$\\
    \hline
    BOTTOMS Stream Temperature  & Bot T & $\text{Bot}_{\text{T}}$\\
    \hline
     Condenser Duty  & Q Cond & $\text{Q}_{\text{cond}}$\\
     \hline
    Reboiler Vapour Flow Rate  & Vep Reb & $\text{Vap}_{\text{Reb}}$\\
    \hline
    Stage 1 (Condenser) Pressure &  P1 & $\text{P}_{1}$\\
    \hline
    Stage 22 (Reboiler) Pressure &  P22 & ${P}_{22}$ \\
    \hline
		\end{tabular}
	\end{center}
	\caption{State Space Variables for Distillation Column Dynamics}
	\label{StateVariables}
\end{table}

These variables hold significance in terms of column requirements as the equations developed can later be used for obtaining a specific extent of separation, quality of product, ensure pressure in the column within safety limits or estimate energy requirements. ODEs in terms of these variables will make these use cases possible.\\
However, due to the presence of trace quantities of chemicals in streams, the equations for those chemicals produced inaccurate results. This can be attributed to the low Signal to Noise Ratio (noise arising from numerical integration) for these variables. To improve the model, the total flow rate of the overhead stream was also included as state variables. The total flow rate is a redundant  variable as it can be estimated as a summation of the individual flow rates. But, since the total flow has a higher SNR, it is expected to produce better results.
\subsubsubsection{\textbf{Candidate Functions}}
The variables were first mean shifted and auto scaled before generating the candidate functions.\\
We used 360 candidate functions of the form,
\begin{align*}
f_i &=\; x_1^{a_1^{(i)}}x_2^{a_2^{(i)}}\dotsm x_{14}^{a_{14}^{(i)}} \quad \quad i = 1,2\hdots k \\
&\text{Where,} \\
&\displaystyle\Sigma_{j=1}^{14} \leq 2\\
-&2 \leq a_j^{(i)} \leq 2\\
& a_j^{(i)} \in \mathbb{Z}
\end{align*}
And 70 candidate functions of the form, $\sin(x_i),\,\cos(x_i),\, ln(|x_i|),\, e^{x_i},\,\sqrt{|x_i|}\; \forall \; i = 1,2\dots14$. These functions were chosen without using any strong understanding of the system to check if the algorithm can work with very little to no domain knowledge.
\subsubsection{Algorithm Execution}
The algorithm was implemented on Python 3.6.5 using the libraries - \textit{pandas}, \textit{numpy}, \textit{sklearn}, \textit{scipy}, \textit{matplotlib} and \textit{itertools}. We used numerical differentiation with total variance regularization method developed in \cite{chartrand2011numdiff} to obtain the derivatives of the variables. The data was split in the ratio 3:1:1 for training, cross validation and testing. T
\subsection{Model Selection Metrics}
Two methods were used to select models. These two methods differ on the final use of the model and navigate a trade-off between accuracy of prediction and interpretability of the obtained equations. The models differed in the value of the L1 norm regularization parameter. Models with low regularization parameter had a higher number of terms and higher training set accuracy than those with high regularization. 
\subsubsection{Cross Validation Accuracy}
\begin{figure*}[h!]
    \centering
    \begin{subfigure}[t]{0.49\textwidth}
        \centering
        \includegraphics[height=1.2in]{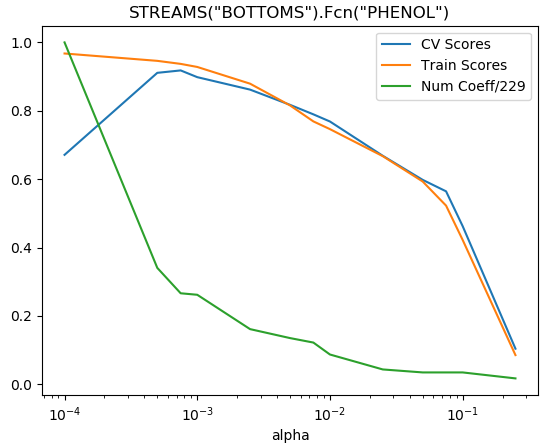}
        \caption{Clear cross-validation peak}
        \label{fig:x8}
    \end{subfigure}
    \begin{subfigure}[t]{0.5\textwidth}
        \centering
        \includegraphics[height=1.2in]{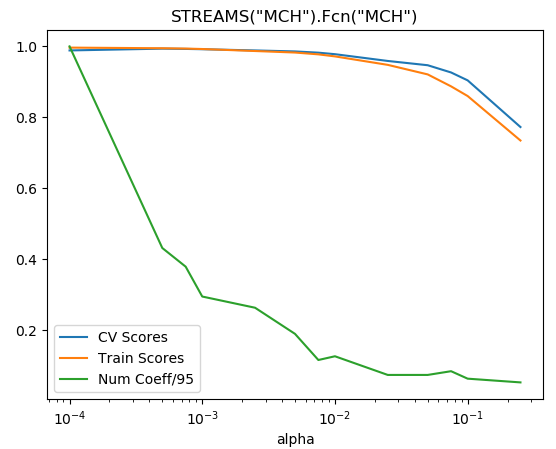}
        \caption{Without a clear cross-validation peak}
        \label{fig:x3}
    \end{subfigure}
    \caption{Cross Validation Model Selection}
\end{figure*}

This selected the model with highest cross validation accuracy. For some variables, cross validation had a clear peak as shown in Fig.\ref{fig:x8}. The accuracy is expected to initially increase with reducing regularization but later reduce due to over fitting. However, some of the selected models had too many terms making it difficult to interpret their physical meaning. Also, some variables like in Fig.\ref{fig:x3} did not exhibit this clear peak characteristic and the accuracy kept increasing with smaller regularization. The implications of this observation are discussed in Section \ref{sec:results}.
\subsubsection{Cross Validation Accuracy with Model Complexity Penalty}
To avoid selecting a lot of terms and to break ties in cases without a clear cross validation peak, a selection score based on model complexity was defined. Based on the score given by Eq.\ref{eq:score}, the model with the highest score was selected.
\begin{align}
    \alpha k-\beta ln\left(R^2_{CV}\right) \label{eq:score}
\end{align}
where $\alpha$ and $\beta$ are weights selected based on inspection of the trade off graphs. $k$ denotes the number of terms in the obtained equation and $R^2_{CV}$ is the cross validation $R^2$ accuracy.
\subsection{Testing}
\label{subsec:testing}
Different testing methods were employed to quantify the goodness of the developed equations for different purposes. We looked at the accuracy of predicting $\bm{\dot{X}(t)}$ given $\bm{X(t)}$ and $\bm{u(t)}$ and the accuracy in predicting $\bm{X(t)}$ from $\bm{u(t)}$ and the initial condition, by integrating the ODEs obtained. The results of these tests along with their interpretations are available in Section \ref{sec:results} and Appendix \ref{app:result}. The methods employed were:
\begin{description}
    \item [Test Data] Tests the accuracy of the developed model on the $20\%$ data selected randomly and excluded from training. This gives an idea about the overfitting and the predictive ability of the model under conditions similar to which the training data was obtained. Low success under this test could indicate overfitting.
    \item [Outside Perturbation Region] This creates a new data set by changing the feed perturbation region and testing the model on this new data. This checks if the model was able to capture the complete dynamics of the model. Low accuracy under this test would indicate incompleteness of the model in terms of missing critical state variables or insufficient candidate functions.
    \item [Long Time Accuracy] In this testing, the dynamical system is run for a longer time (250 hours) than the training time (100 hours) to generate test data. This will help identify long time dynamic effects or time based evolution of the system which could have been missed by the algorithm.
    \item [Similar System Structural Comparison] 4 additional systems were created as mentioned in Table. \ref{tab:op_cond} by altering the operating conditions. The model was trained on these 4 systems. The structure of the equations obtained were compared across these 5 systems for similar terms (only for the presence or absence of terms and not for the similarity of regression coefficients). If the algorithm is able to extract the entire dynamics of the system, irrespective of the operating condition, the equation would contain the same terms and differ only in the parameter values.
\end{description}
\section{Results and Analysis}
\label{sec:results}
\subsection{Derivative Predictions}
\label{subsec:accuracy}
\begin{figure}[H]
\begin{subfigure}{.5\textwidth}
  \centering
  \includegraphics[width=.8\linewidth]{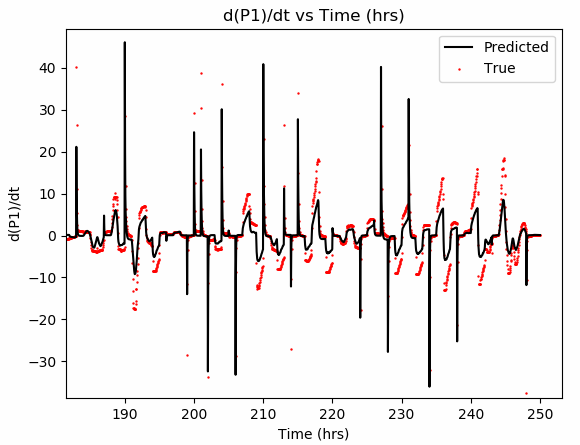}
  \caption{Condenser Pressure Derivative vs Time - Good Predictions}
\end{subfigure}
\begin{subfigure}{.5\textwidth}
  \centering
  \includegraphics[width=.8\linewidth]{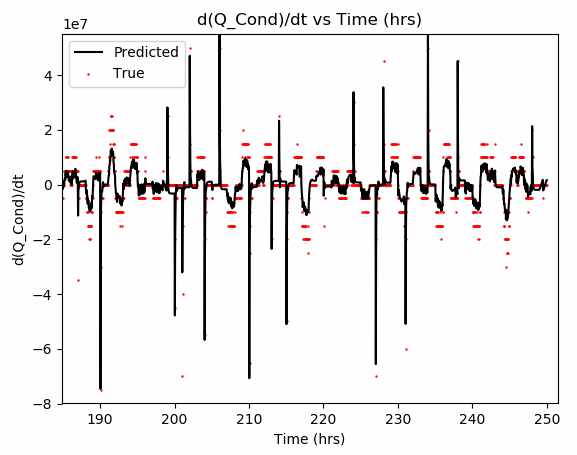}
  \caption{Reboiler Duty Derivative vs Time - Good Predictions}
\end{subfigure}
\caption{Long Time Data Set}
\label{fig:dydx_vs_t_long}
\end{figure}

\begin{figure}[H]
\begin{subfigure}{.5\textwidth}
  \centering
  \includegraphics[width=.8\linewidth]{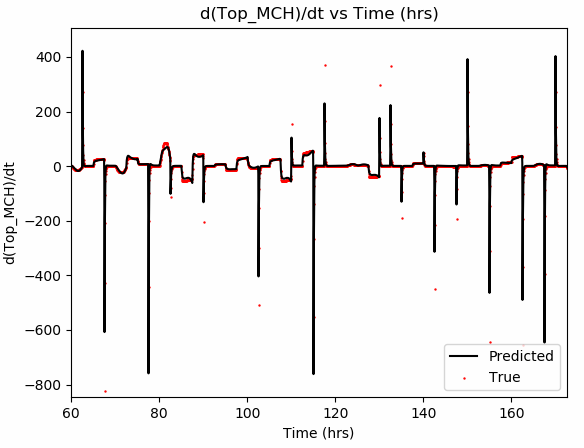}
  \caption{Outside Training Perturbation - Good Predictions}
  \label{fig:out_good}
\end{subfigure}
\begin{subfigure}{.5\textwidth}
  \centering
  \includegraphics[width=.8\linewidth]{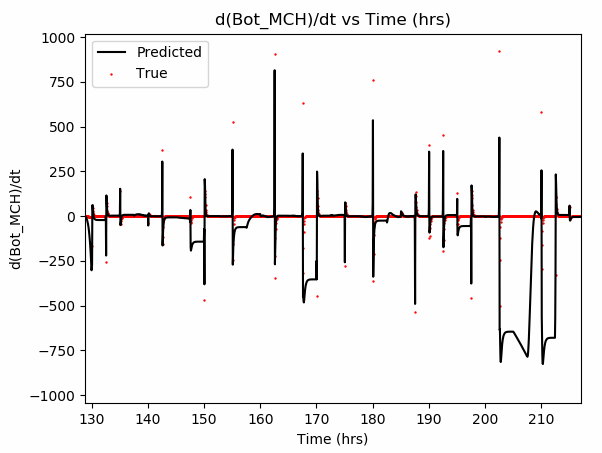}
  \caption{Outside Training Perturbation - Poor Predictions}
  \label{img:out_bad}
\end{subfigure}
\caption{Outside Training Perturbation Data Set}
\label{fig:dydx_vs_t_out}
\end{figure}

The model was trained on the four systems mentioned in Table. \ref{tab:op_cond}. The developed equations were used to predict $\bm{\dot{X}(t)}$ from $\bm{X(t)}$ and the input for the test data. These results along with the sparsity of the models given by the number of terms N are given in Table. \ref{tab:test_accuracy1} and \ref{tab:test_accuracy2}. The training and testing were done for 2 values of $\alpha$ corresponding to low regularization and high regularization. 
\begin{longtable}[c]{|l|
>{\columncolor[HTML]{EFEFEF}}r |
>{\columncolor[HTML]{EFEFEF}}r |
>{\columncolor[HTML]{EFEFEF}}r |
>{\columncolor[HTML]{EFEFEF}}r |
>{\columncolor[HTML]{EFEFEF}}r |
>{\columncolor[HTML]{EFEFEF}}r |r|r|r|r|r|r|}
\caption{Training and Test $R^2$ values for the 4 systems}
\label{tab:test_accuracy1}\\
\hline
\multicolumn{1}{|c|}{} & \multicolumn{6}{c|}{\cellcolor[HTML]{EFEFEF}\textbf{Basic}} & \multicolumn{6}{c|}{\textbf{MCH400}} \\ \cline{2-13} 
\multicolumn{1}{|c|}{} & \multicolumn{3}{c|}{\cellcolor[HTML]{EFEFEF}\textbf{\begin{tabular}[c]{@{}c@{}}Low\\   Regularization\end{tabular}}} & \multicolumn{3}{c|}{\cellcolor[HTML]{EFEFEF}\textbf{High Regularization}} & \multicolumn{3}{c|}{\textbf{\begin{tabular}[c]{@{}c@{}}Low\\   Regularization\end{tabular}}} & \multicolumn{3}{c|}{\textbf{High Regularization}} \\ \cline{2-13} 
\multicolumn{1}{|c|}{\multirow{-3}{*}{\textbf{Variable}}} & \multicolumn{1}{c|}{\cellcolor[HTML]{EFEFEF}\textbf{Train}} & \multicolumn{1}{c|}{\cellcolor[HTML]{EFEFEF}\textbf{Test}} & \multicolumn{1}{c|}{\cellcolor[HTML]{EFEFEF}\textbf{N}} & \multicolumn{1}{c|}{\cellcolor[HTML]{EFEFEF}\textbf{Train}} & \multicolumn{1}{c|}{\cellcolor[HTML]{EFEFEF}\textbf{Test}} & \multicolumn{1}{c|}{\cellcolor[HTML]{EFEFEF}\textbf{N}} & \multicolumn{1}{c|}{\textbf{Train}} & \multicolumn{1}{c|}{\textbf{Test}} & \multicolumn{1}{c|}{\textbf{N}} & \multicolumn{1}{c|}{\textbf{Train}} & \multicolumn{1}{c|}{\textbf{Test}} & \multicolumn{1}{c|}{\textbf{N}} \\ \hline
\endhead
\textbf{Top F} & {\color[HTML]{333333} 0.99} & 0.986 & 25 & 0.979 & 0.979 & 14 & 0.959 & 0.942 & 39 & 0.93 & 0.897 & 19 \\ \hline
\textbf{Top T} & 0.99 & 0.985 & 27 & 0.978 & 0.977 & 15 & 0.965 & 0.943 & 36 & 0.933 & 0.894 & 18 \\ \hline
\textbf{Top MCH} & 0.99 & 0.986 & 28 & 0.979 & 0.979 & 14 & 0.96 & 0.943 & 39 & 0.93 & 0.898 & 19 \\ \hline
\textbf{Top Ph} & 0.282 & 0.315 & 6 & 0.282 & 0.315 & 6 & 0.435 & 0.389 & 50 & 0.377 & 0.358 & 40 \\ \hline
\textbf{Top Tol} & 0.962 & 0.974 & 25 & 0.959 & 0.969 & 18 & 0.954 & 0.924 & 41 & 0.866 & 0.712 & 9 \\ \hline
\textbf{Bot T} & 0.972 & 0.966 & 50 & 0.854 & 0.835 & 19 & 0.875 & 0.774 & 60 & 0.748 & 0.669 & 23 \\ \hline
\textbf{Bot MCH} & 0.975 & 0.977 & 45 & 0.827 & 0.815 & 22 & 0.856 & 0.772 & 63 & 0.698 & 0.58 & 22 \\ \hline
\textbf{Bot Ph} & 0.904 & 0.871 & 57 & 0.718 & 0.632 & 23 & 0.795 & 0.755 & 78 & 0.544 & 0.495 & 23 \\ \hline
\textbf{Bot Tol} & 0.873 & 0.722 & 50 & 0.723 & 0.5 & 15 & 0.77 & 0.769 & 63 & 0.58 & 0.53 & 21 \\ \hline
\textbf{Cond Q} & 0.972 & 0.956 & 36 & 0.958 & 0.948 & 14 & 0.955 & 0.926 & 50 & 0.909 & 0.858 & 20 \\ \hline
\textbf{Vap Reb} & 0.914 & 0.866 & 37 & 0.844 & 0.783 & 20 & 0.861 & 0.822 & 64 & 0.686 & 0.651 & 20 \\ \hline
\textbf{P1} & 0.976 & 0.96 & 31 & 0.963 & 0.939 & 14 & 0.952 & 0.908 & 32 & 0.927 & 0.862 & 19 \\ \hline
\textbf{P22} & 0.965 & 0.948 & 30 & 0.947 & 0.923 & 16 & 0.932 & 0.902 & 52 & 0.873 & 0.812 & 18 \\ \hline
\end{longtable}
\begin{longtable}[c]{|l|
>{\columncolor[HTML]{EFEFEF}}r |
>{\columncolor[HTML]{EFEFEF}}r |
>{\columncolor[HTML]{EFEFEF}}r |
>{\columncolor[HTML]{EFEFEF}}r |
>{\columncolor[HTML]{EFEFEF}}r |
>{\columncolor[HTML]{EFEFEF}}r |r|r|r|r|r|r|}
\caption{Training and Test $R^2$ values for the 4 systems}
\label{tab:test_accuracy2}\\
\hline
\multicolumn{1}{|c|}{} & \multicolumn{6}{c|}{\cellcolor[HTML]{EFEFEF}\textbf{T400}} & \multicolumn{6}{c|}{\textbf{RR6}} \\ \cline{2-13} 
\multicolumn{1}{|c|}{} & \multicolumn{3}{c|}{\cellcolor[HTML]{EFEFEF}\textbf{\begin{tabular}[c]{@{}c@{}}Low\\   Regularization\end{tabular}}} & \multicolumn{3}{c|}{\cellcolor[HTML]{EFEFEF}\textbf{High Regularization}} & \multicolumn{3}{c|}{\textbf{\begin{tabular}[c]{@{}c@{}}Low\\   Regularization\end{tabular}}} & \multicolumn{3}{c|}{\textbf{High Regularization}} \\ \cline{2-13} 
\multicolumn{1}{|c|}{\multirow{-3}{*}{\textbf{Variable}}} & \multicolumn{1}{c|}{\cellcolor[HTML]{EFEFEF}\textbf{Train}} & \multicolumn{1}{c|}{\cellcolor[HTML]{EFEFEF}\textbf{Test}} & \multicolumn{1}{c|}{\cellcolor[HTML]{EFEFEF}\textbf{N}} & \multicolumn{1}{c|}{\cellcolor[HTML]{EFEFEF}\textbf{Train}} & \multicolumn{1}{c|}{\cellcolor[HTML]{EFEFEF}\textbf{Test}} & \multicolumn{1}{c|}{\cellcolor[HTML]{EFEFEF}\textbf{N}} & \multicolumn{1}{c|}{\textbf{Train}} & \multicolumn{1}{c|}{\textbf{Test}} & \multicolumn{1}{c|}{\textbf{N}} & \multicolumn{1}{c|}{\textbf{Train}} & \multicolumn{1}{c|}{\textbf{Test}} & \multicolumn{1}{c|}{\textbf{N}} \\ \hline
\endhead
\textbf{Top F} & 0.983 & 0.957 & 53 & 0.944 & 0.921 & 21 & 0.989 & 0.963 & 37 & 0.97 & 0.955 & 16 \\ \hline
\textbf{Top T} & 0.985 & 0.981 & 52 & 0.951 & 0.951 & 19 & 0.989 & 0.966 & 40 & 0.964 & 0.949 & 16 \\ \hline
\textbf{Top MCH} & 0.974 & 0.963 & 39 & 0.942 & 0.926 & 20 & 0.989 & 0.963 & 36 & 0.97 & 0.955 & 16 \\ \hline
\textbf{Top Ph} & 0.624 & 0.605 & 34 & 0.625 & 0.605 & 34 & 0.562 & 0.565 & 43 & 0.511 & 0.554 & 30 \\ \hline
\textbf{Top Tol} & 0.97 & 0.609 & 33 & 0.933 & 0.704 & 20 & 0.892 & 0.826 & 8 & 0.892 & 0.826 & 8 \\ \hline
\textbf{Bot T} & 0.878 & 0.734 & 70 & 0.733 & 0.696 & 27 & 0.98 & 0.889 & 41 & 0.892 & 0.491 & 18 \\ \hline
\textbf{Bot MCH} & 0.832 & 0.7 & 67 & 0.649 & 0.49 & 24 & 0.863 & 0.794 & 48 & 0.759 & 0.66 & 31 \\ \hline
\textbf{Bot Ph} & 0.792 & 0.493 & 87 & 0.792 & 0.493 & 87 & 0.832 & 0.762 & 48 & 0.703 & 0.504 & 23 \\ \hline
\textbf{Bot Tol} & 0.803 & 0.389 & 88 & 0.803 & 0.389 & 88 & 0.952 & 0.866 & 30 & 0.923 & 0.89 & 14 \\ \hline
\textbf{Cond Q} & 0.969 & 0.964 & 50 & 0.935 & 0.93 & 18 & 0.966 & 0.926 & 50 & 0.93 & 0.904 & 15 \\ \hline
\textbf{Vap Reb} & 0.889 & 0.875 & 67 & 0.758 & 0.864 & 38 & 0.91 & 0.843 & 51 & 0.822 & 0.677 & 21 \\ \hline
\textbf{P1} & 0.976 & 0.934 & 53 & 0.939 & 0.901 & 25 & 0.982 & 0.97 & 43 & 0.968 & 0.946 & 20 \\ \hline
\textbf{P22} & 0.958 & 0.935 & 51 & 0.91 & 0.908 & 21 & 0.977 & 0.948 & 38 & 0.958 & 0.906 & 21 \\ \hline
\end{longtable}
We find that the model is able to predict $\bm{\dot{X}(t)}$ with a reasonable accuracy from $\bm{X(t)}$ and $u(t)$. Reducing the regularization increases the accuracy in the test data. This trend is seen across variables and till very small regularization parameter values. This indicates that we are unable to capture enough information from the data using the provided candidate functions and number of terms. This could either indicate insufficient candidate function and state variables or absence of a low dimensional function space representation for the system. Ways to analyze and possibly overcome this are discussed in Section. \ref{sec:discuss}.\\~\\
System 1 was also tested on two other simulations (one run for a longer time and the other outside the training perturbation region) explained in Section. \ref{subsec:testing}. The results for these two tests are in Table. \ref{tab:long_out}. Sample result plots of $\bm{\dot{X}(t)}$ vs $t$ for Prediction vs True Values for these tests are in Fig. \ref{fig:dydx_vs_t_long} and \ref{fig:dydx_vs_t_out}.
\begin{longtable}[c]{|l|
>{\columncolor[HTML]{EFEFEF}}r |
>{\columncolor[HTML]{EFEFEF}}r |r|r|}
\caption{Long Time and Testing Outside the Training Perturbation Region}
\label{tab:long_out}\\
\hline
\multicolumn{1}{|c|}{} & \multicolumn{2}{c|}{\cellcolor[HTML]{EFEFEF}\textbf{Long Time}} & \multicolumn{2}{c|}{\textbf{Outside Training}} \\ \cline{2-5} 
\multicolumn{1}{|c|}{\multirow{-2}{*}{\textbf{Variable}}} & \multicolumn{1}{c|}{\cellcolor[HTML]{EFEFEF}\textbf{Low $\alpha$}} & \multicolumn{1}{c|}{\cellcolor[HTML]{EFEFEF}\textbf{High $\alpha$}} & \multicolumn{1}{c|}{\textbf{Low $\alpha$}} & \multicolumn{1}{c|}{\textbf{High $\alpha$}} \\ \hline
\endfirsthead
\endhead
\textbf{Top F} & 0.948 & 0.945 & 0.778 & 0.803 \\ \hline
\textbf{Top T} & 0.953 & 0.944 & 0.704 & 0.817 \\ \hline
\textbf{Top MCH} & 0.951 & 0.946 & 0.819 & 0.799 \\ \hline
\textbf{Top Ph} & 0.198 & 0.198 & -0.588 & -0.588 \\ \hline
\textbf{Top Tol} & 0.516 & 0.523 & 0 & 0.19 \\ \hline
\textbf{Bot T} & 0.95 & 0.813 & -4.94 & -0.477 \\ \hline
\textbf{Bot MCH} & 0.931 & 0.76 & 0 & 0.13 \\ \hline
\textbf{Bot Ph} & 0.786 & 0.585 & -9.034 & -9.034 \\ \hline
\textbf{Bot Tol} & 0.75 & 0.514 & -30.744 & -30.744 \\ \hline
\textbf{Cond Q} & 0.949 & 0.922 & 0.626 & 0.789 \\ \hline
\textbf{Vap Reb} & 0.852 & 0.78 & -0.498 & -1.424 \\ \hline
\textbf{P1} & 0.868 & 0.86 & 0.24 & 0.783 \\ \hline
\textbf{P22} & 0.851 & 0.877 & 0 & 0.726 \\ \hline
\end{longtable}
We see that the model performs very well in the Long Time data. This cements the fact that the evolution of the system with time (if present) has been captured. If this weren't the case the model performance would have deteriorated with longer tests.
However the performance is sub par in region outside the training perturbation for most of the states. Also, with higher regularization, the model marginally improves as opposed to all the previous observations where the model kept getting better on the test set with decreasing regularization. This indicates that the available variables and candidate functions are over fitting not the training data but the state of the system in the training region. This can be resolved by including new state variables which will make the model obtained invariant to the training perturbation region. 
In Fig. \ref{img:out_bad} we can see a variable for which the algorithm has poor results. The model predicts peaks in regions of steady operation. This could be because a limiting variable which dictates the dynamics in this region has been missed. So, the system does not exhibit significant dynamics here, while the model predicts dynamics. However, in Fig. \ref{fig:out_good} the prediction closely follows the model. By using these two equations for the same component in different outlet streams we can try to understand which states might be missed. By incorporating the missed states we can iteratively improve the model.
\subsection{ODE Structure Comparison}
The ODEs obtained for the 4 systems were compared with each other for similarity in the terms selected. The number of such similar terms for two levels of regularization along with the total number of terms is provided in Appendix \ref{app:ode_common}. Appendix \ref{app:ode_common} also has a list of the terms with most repetitions cross the 4 systems tested. These results can be interpreted as dynamic equivalent of sensitivity analysis in steady systems. If the complete dynamics had been captured, most of the terms in the ODEs would have been repeated across the systems. However, this was not observed. Hardly $10\%$ of the total terms were common across 3 systems where the feed compositions were altered. This could mean that we have not completely described our system with the current set of states. We need to look for variables which are crucial in deciding the dynamics by performing sensitivity analyses on the operating conditions too.\\
However, by reducing regularization we notice that the fraction of terms retained across the systems either increases or remains the same in most cases. This indicates that by increasing the number of candidate functions selected, they are able to explain the model better, even if only by a small increment. This result correlates with the prediction accuracy explained in Section. \ref{subsec:accuracy} which kept improving with smaller regularization. We also find the same terms repeating across all 4 systems more commonly. The system with a different Reflux Ratio (which is the only column specification varied) had the no common terms with the other systems under high regularization but had an increasing number of common terms under low regularization. This could further indicate that the system might not be truly sparse in function space, highlighting the possible limitations of using SINDy in identifying the complex dynamics of unknown system without some knowledge about functional space that may govern the dynamics of these systems.

A similar analysis was carried out between the training set and the test set with phenol feed outside the training perturbation region. The results of this analysis are listed in Table \ref{tab:ode_comp_out}
\begin{longtable}[c]{|l|
>{\columncolor[HTML]{EFEFEF}}r |
>{\columncolor[HTML]{EFEFEF}}r |r|r|}
\caption{Structural Similarity of ODEs}
\label{tab:ode_comp_out}\\
\hline
\multicolumn{1}{|c|}{} & \multicolumn{2}{c|}{\cellcolor[HTML]{EFEFEF}\textbf{Low $\alpha$}} & \multicolumn{2}{c|}{\textbf{High $\alpha$}} \\ \cline{2-5} 
\multicolumn{1}{|c|}{\multirow{-2}{*}{\textbf{Variable}}} & \multicolumn{1}{c|}{\cellcolor[HTML]{EFEFEF}\textbf{Common}} & \multicolumn{1}{c|}{\cellcolor[HTML]{EFEFEF}\textbf{Total}} & \multicolumn{1}{c|}{\textbf{Common}} & \multicolumn{1}{c|}{\textbf{Total}} \\ \hline
\endhead
\textbf{Top F} & 5 & 23 & 5 & 14 \\ \hline
\textbf{Top T} & 4 & 24 & 1 & 3 \\ \hline
\textbf{Top MCH} & 7 & 28 & 4 & 14 \\ \hline
\textbf{Top Ph} & 1 & 6 & 1 & 6 \\ \hline
\textbf{Top Tol} & 13 & 28 & 5 & 18 \\ \hline
\textbf{Bot T} & 18 & 39 & 5 & 16 \\ \hline
\textbf{Bot MCH} & 21 & 39 & 3 & 18 \\ \hline
\textbf{Bot Ph} & 19 & 35 & 6 & 13 \\ \hline
\textbf{Bot Tol} & 21 & 37 & 4 & 12 \\ \hline
\textbf{Cond Q} & 13 & 36 & 3 & 14 \\ \hline
\textbf{Vap Reb} & 17 & 35 & 5 & 17 \\ \hline
\textbf{P1} & 5 & 30 & 3 & 9 \\ \hline
\textbf{P22} & 17 & 34 & 3 & 11 \\ \hline
\end{longtable}
We see that lowering regularization in most of the cases decreases the fraction of common terms. However in the other systems tested inside the perturbation region, this was not the case. This along with the interpretations of Table. \ref{tab:long_out} further confirm the fact that some crucial state variables or functional forms are being missed. Even though the prediction accuracy is high, if the true mechanisms are captured for these complex systems, same terms should appear in the governing equations. However, if the aim of the work is to obtain simpler equations that can capture the non-linear dynamics, the algorithm performs well. But, in order to understand the true physical mechanisms, the SINDy algorithm perhaps need to be provided with functional forms determined by domain experts. As was done in the reaction kinetics identification \cite{hoffmann2019}, the authors provided functional form determined by ``law of mass action" which is a known physical law that drives rate kinetics and mechanisms. To improve on the distillation column differential equation identification, such knowledge about relationship between top and bottom feed, temperature and pressure need to be used to construct appropriate functions. This is challenging for the distillation column system because there are several heuristic based equations that are used in design of the seperation system. Our future work will address this need of converting these complex design equations that govern non-equilibrium system in distillation column to appropriate functional forms to be used in extracting the governing differential equations.

\subsection{Structure of ODEs and Physical Interpretations}
The Structure of the ODE obtained for the basic system under high regularization is available in Appendix \ref{app:ode}. While these ODEs are very complex to interprete, it is still a win for representing the dynamics of this system using one equation for each state variable as compared to over 1000 complex equations that relate the dynamics of system. However, there was no direct interpretation of most of the terms in physical sense. Some of the terms such as $sin(Top_{Tol})$ which represents sin of Toluene concentration in Top flow is physically not interpretable. Some of the commonly recurring terms that we found physically relevant were : $Conc^{2}$ which basically meant that second order terms in concentration were found relevant for controlling the dynamics. The only feasible interpretation can be that diffusion of  two components or cross diffusion is driving the dynamics. This can be because of fick's law of diffusion acting on both the component involved. For example one terms in Appendix is $Bot_{MCH}/Bot_{Tol}$ which is the ratio of concentration of MCH and Toluene in bottom flow. Apperance of this term in the equation driving the MCH in top stream denotes some relationship between diffusivity difference of MCH and Toluene in the extracting component Phenol. While the form of equation is surprising because the functional form did not give the fick's law of diffusion which actually needs concentration variation rather than just concentration, the appearance of this term provides some hope of these data driven approaches to learn about the governing mechanisms of dynamics in complex systems. Another term that we relate to a physical law is ratio of concentration and Pressure such as the term $Bot_{Tol}/P_{22}$. This term represents concentration of Toluene in bottom feed and pressure of the last plate. We related this term to the Henry's law which relates concentration of a solute in liquid phase to the partial pressure of the solute in gas phase. This term probably represents the relationship that the dynamics of concentration in bottom stream for toluene is related to the pressure on plates where the component may exist in vapor phase. The gas-liquid mass transfer in these systems are interconnected and complex, hence it is difficult to pin-point one single mechanism driving dynamics. However, it is encouraging to see some functional forms that may be related to physical laws being picked up in these equations. In order to be able to identify the laws, more complex rules for application of machine learning in complex engineered systems must be developed.


\section{Discussions}
\label{sec:discuss}
In this work our goal was to apply a machine learning approach on data generated from mechanistic model for distillation column to test the hypothesis of identifying governing physical laws for dynamics of the system. We began with the approach of Sparse Identification of Non-Linear Dynamics (SINDy) because of it's ability to give white box models and allow interpretation of terms that drive the dynamics. We tested both for accuracy and also the changes in structure of equations obtained under different design consideration of the system. We picked a distillation column because of it's ubiquitious role in chemical engineering world from petrochemical industries to biomass refining. The results for prediction on the test data generated from mechanistic models were very encouraging with most variables showing more than 80 \% accuracy. Outside the perturbation range, the equations did not perform very well which may be because of the change in dynamic regime. If the training data set only captured a particular dynamic regime, it cannot capture the dynamics in a different regime. However, this is still an un-resolved question from mechanistic perspective, that if DEs capture truly physical mechanisms this should provide insights into impending regime change as well. From physical interpretation perspective of the equations obtained, it was encouraging to see terms such as $Concentration^{2}$ and ratio of concentration with pressure. The prior can be related to fick's law of diffusion for two components in the column whereas the later can be related to the Henry's law controlling the solubility of the components in the mixture controlled by pressure at different plates in the column. One interesting finding from extracting these DEs is the simplified relationship that was obtained between component flow rate in top stream to the component flow rates in the bottom flow rate along with the pressure of last plate. In actual distillation column design, there is a mass balance equation solved for each plate that finally relates the component concentration in top stream to the bottom stream. Use of this one simplified equation captures this whole dynamics. We think this is the strength of machine learning approach. Based on the accuracy of prediction within certain time steps, a moving time window to train the model would be more appropriate. We expect that this can be used in better control systems design because the method can capture the non-linear dynamics much better and the need of linearization as prevalent in traditional control design may be relaxed. At the end, novel machine learning advancement is opening up new avenues of looking at complex engineered systems where traditional first principle method of extracting governing equations may fail. However, we are still a long way to go. A greater cross communication between engineering and data science would be required to achieve breakthroughs in limitations of engineering dynamical studies using machine learning approach. Both fields must inform each other for overcoming the limitations in algorithms as well.

\bibliographystyle{unsrtnat}  
\bibliography{references}
\appendix
\appendixpage
\addappheadtotoc
\label{app:result}
\section{ODEs Obtained}
\label{app:ode}
The ODEs obtained for the basic system under high regularization are reported here.\\
$\dot{\text{Top}_F} = 0.4093\text{V}_{\text{Reb}} \text{P} _{22}^{-1} - 0.01251\text{Bot}_{\text{Tol}} \text{Phenol} - 0.697\text{Bot}_{\text{Tol}} \text{P} _{22}^{-1} - 1.284\text{Bot}_{\text{MCH}} \text{Bot}_{\text{Tol}}^{-1} - 0.3442\text{Bot}_{\text{MCH}}^2 + 0.0516\text{Bot}_T \text{Bot}_{\text{Tol}}^{-1} + 0.04887\text{Top}_{\text{Tol}}^{-1} \text{Phenol}^{-1} + 0.1724\text{Top}_{\text{Tol}} \text{Bot}_{\text{MCH}} + 0.02372\text{Top}_{\text{Ph}}^{-1} \text{Phenol}^{-1} + 0.04604\text{Top}_{\text{Ph}}^{-1} \text{P}_1 - 0.09515\text{Top}_F^{-1} \text{Bot}_{\text{Tol}} - 0.03098sin(\text{Top}_{\text{Tol}}) - 0.0105sin(\text{Bot}_{\text{MCH}}) - 0.006331cos(\text{Top}_{\text{Tol}})$\\~\\
  $\dot{\text{Top}_T} = 0.3748\text{V}_{\text{Reb}} \text{P} _{22}^{-1} - 0.01408\text{Bot}_{\text{Tol}} \text{Phenol} - 0.8135\text{Bot}_{\text{Tol}} \text{P} _{22}^{-1} - 0.002575\text{Bot}_{\text{MCH}}^{-1} \text{Bot}_{\text{Tol}} - 1.386\text{Bot}_{\text{MCH}} \text{Bot}_{\text{Tol}}^{-1} - 0.2882\text{Bot}_{\text{MCH}}^2 + 0.001331\text{Bot}_T \text{P} _{22}^{-1} + 0.06055\text{Bot}_T \text{Bot}_{\text{Tol}}^{-1} + 0.09324\text{Top}_{\text{Tol}}^{-1} \text{Phenol}^{-1} + 0.000515\text{Top}_{\text{Tol}}^{-1} \text{P}_1 + 0.1343\text{Top}_{\text{Tol}} \text{Bot}_{\text{MCH}} + 0.07353\text{Top}_{\text{Ph}}^{-1} \text{P}_1 - 0.003506\text{Top}_{\text{Ph}} \text{Bot}_{\text{MCH}}^{-1} - 0.03555sin(\text{Top}_{\text{Tol}}) - 0.01057sin(\text{Bot}_{\text{MCH}})$\\~\\
  $\dot{\text{Top}_{\text{MCH}}} = 0.409\text{V}_{\text{Reb}} \text{P} _{22}^{-1} - 0.01188\text{Bot}_{\text{Tol}} \text{Phenol} - 0.6272\text{Bot}_{\text{Tol}} \text{P} _{22}^{-1} - 1.278\text{Bot}_{\text{MCH}} \text{Bot}_{\text{Tol}}^{-1} - 0.3459\text{Bot}_{\text{MCH}}^2 + 0.05691\text{Bot}_T \text{Bot}_{\text{Tol}}^{-1} + 0.04383\text{Top}_{\text{Tol}}^{-1} \text{Phenol}^{-1} + 0.1908\text{Top}_{\text{Tol}} \text{Bot}_{\text{MCH}} + 0.03016\text{Top}_{\text{Ph}}^{-1} \text{Phenol}^{-1} + 0.06078\text{Top}_{\text{Ph}}^{-1} \text{P}_1 - 0.1535\text{Top}_F^{-1} \text{Bot}_{\text{Tol}} - 0.03093sin(\text{Top}_{\text{Tol}}) - 0.01047sin(\text{Bot}_{\text{MCH}}) - 0.002557cos(\text{Top}_{\text{Tol}})$\\~\\
  $\dot{\text{Top}_{\text{Ph}}} = -0.1884\text{Bot}_{\text{MCH}} \text{V}_{\text{Reb}} + 0.1618\text{Bot}_{\text{MCH}}^{-1} \text{Bot}_{\text{Tol}} - 0.05413\text{Top}_{\text{Tol}} \text{Bot}_{\text{MCH}} - 0.3422\text{Top}_{\text{Tol}}^2 + 0.02866\text{Top}_T \text{V}_{\text{Reb}}^{-1} + 0.01056sin(\text{Top}_{\text{Tol}})$\\~\\
  $\dot{\text{Top}_{\text{Tol}}} = 0.7169\text{Phenol}^{-2} - 0.009678\text{P}_1^{-1} \text{Phenol}^{-1} + 0.01239\text{Bot}_{\text{Tol}} \text{Phenol} + 0.005147\text{Bot}_{\text{Tol}}^{-2} + 0.05915\text{Bot}_{\text{MCH}} \text{Phenol} - 0.1658\text{Bot}_{\text{MCH}}^{-1} \text{Phenol} - 0.01436\text{Bot}_{\text{MCH}} \text{Phenol}^{-1} - 0.06051\text{Bot}_{\text{MCH}}^{-2} - 0.6682\text{Top}_{\text{Tol}}\text{Bot}_{\text{MCH}} + 0.5816\text{Top}_{\text{Tol}} \text{Bot}_{\text{MCH}}^{-1} + 0.02369\text{Top}_T \text{V}_{\text{Reb}}^{-1} + 0.04482exp(\text{Phenol}) + 0.1242sin(\text{Top}_{\text{Tol}}) - 0.05113sin(\text{P}_1) + 0.00829sin(\text{P} _{22}) + 0.8209cos(\text{Top}_{\text{Tol}}) + 0.001593cos(\text{Bot}_{\text{Tol}}) + 0.002661cos(\text{Phenol})$\\~\\
  $\dot{\text{Bot}_T} = -0.0599\text{Bot}_{\text{Tol}} \text{Phenol} + 1.207\text{Bot}_{\text{Tol}}^{-1} \text{P} _{22} - 0.7216\text{Bot}_{\text{Tol}} \text{P}_1^{-1} - 0.3439\text{Bot}_{\text{MCH}}^{-1} \text{Phenol}^{-1} - 0.1131\text{Bot}_{\text{MCH}}^{-1} \text{V}_{\text{Reb}}^{-1} - 0.317\text{Bot}_{\text{MCH}}^{-1} \text{Bot}_{\text{Tol}} - 0.754\text{Bot}_{\text{MCH}}^2 + 0.4488\text{Top}_{\text{Tol}} \text{Phenol} - 0.2163\text{Top}_{\text{Tol}}^{-1} \text{Bot}_{\text{MCH}}^{-1} + 0.09818\text{Top}_{\text{Ph}}^{-1} \text{Bot}_{\text{Tol}}^{-1} + 0.1796\text{Top}_{\text{Ph}}^{-1} \text{Top}_{\text{Tol}} + 0.2743sin(\text{Top}_{\text{Tol}}) - 0.0002233sin(Q_\text{cond}) + 0.02981sin(\text{P} _{22}) + 0.0004682sin(\text{Phenol}) - 0.0829cos(\text{Top}_{\text{Tol}}) - 0.00542cos(\text{Bot}_{\text{Tol}}) - 0.001697cos(\text{V}_{\text{Reb}}) + 0.05604cos(\text{P} _{22})$\\~\\
  $\dot{\text{Bot}_{\text{MCH}}} = 0.02734\text{P}_1^{-1} \text{Phenol}^{-1} + 0.07226\text{Bot}_{\text{Tol}} \text{Phenol} + 1.079\text{Bot}_{\text{Tol}} \text{P}_1^{-1} + 0.4097\text{Bot}_{\text{MCH}}^{-1} \text{V}_{\text{Reb}}^{-1} + 0.6738\text{Bot}_{\text{MCH}} \text{Bot}_{\text{Tol}} + 0.1128\text{Bot}_{\text{MCH}}^{-1} \text{Bot}_{\text{Tol}} - 1.776\text{Bot}_{\text{MCH}} \text{Bot}_{\text{Tol}}^{-1} + 1.245\text{Bot}_{\text{MCH}}^2 - 0.4121\text{Top}_{\text{Tol}} \text{Phenol} + 0.2065\text{Top}_{\text{Tol}}^{-1} \text{Bot}_{\text{Tol}} - 0.08311\text{Top}_{\text{Tol}} \text{Bot}_{\text{MCH}} - 0.5822\text{Top}_{\text{Ph}} \text{Bot}_{\text{MCH}} + 0.1525\text{Top}_{\text{Ph}} \text{Top}_{\text{Tol}}^{-1} - 0.01886\text{Top}_{\text{MCH}} \text{Bot}_{\text{Tol}}^{-1} + 0.05897\text{Top}_T \text{V}_{\text{Reb}}^{-1} - 0.08836\text{Top}_F \text{Bot}_{\text{Tol}}^{-1} - 0.2456sin(\text{Top}_{\text{Tol}}) - 0.1025sin(\text{P}_1) - 0.003293sin(\text{Phenol}) + 0.08151cos(\text{Top}_{\text{Tol}}) + 0.01463cos(\text{Bot}_{\text{Tol}}) + 0.0008987cos(\text{V}_{\text{Reb}})$\\~\\
  $\dot{\text{Bot}_{\text{Ph}}} = -0.04869\text{Bot}_{\text{Tol}} \text{Phenol} + 0.2687\text{Bot}_{\text{Tol}}^{-1} \text{P} _{22} + 0.4737\text{Bot}_{\text{Tol}}^{-1} \text{P}_1 - 0.5239\text{Bot}_{\text{Tol}} \text{P}_1^{-1} + 0.1406\text{Bot}_{\text{Tol}}^2 - 0.4105\text{Bot}_{\text{MCH}}^{-1} \text{Phenol}^{-1} - 0.09249\text{Bot}_{\text{MCH}} \text{Bot}_{\text{Tol}} - 0.6061\text{Bot}_{\text{MCH}}^{-1} \text{Bot}_{\text{Tol}} + 1.515\text{Bot}_{\text{MCH}} \text{Bot}_{\text{Tol}}^{-1} - 0.7218\text{Bot}_{\text{MCH}}^2 + 0.516\text{Top}_{\text{Tol}} \text{Phenol} - 0.2959\text{Top}_{\text{Tol}}^{-1} \text{Bot}_{\text{Tol}} - 0.001671\text{Top}_{\text{Tol}}^{-1} \text{Bot}_{\text{MCH}}^{-1} + 0.5482\text{Top}_{\text{Ph}}^{-1} \text{Top}_{\text{Tol}} - 0.07217\text{Top}_T^{-1} \text{V}_{\text{Reb}} + 0.1484\text{Top}_F \text{Bot}_{\text{Tol}}^{-1} + 0.5772sin(\text{Top}_{\text{Tol}}) - 0.004738sin(\text{Bot}_{\text{MCH}}) - 0.006349sin(Q_\text{cond}) + 0.01085sin(\text{Phenol}) - 0.02076cos(\text{Top}_{\text{Tol}}) - 0.01243cos(\text{Bot}_{\text{Tol}}) + 0.1466cos(\text{P} _{22}) $\\~\\
  $\dot{\text{Bot}_{\text{Tol}}} = -0.0716\text{Bot}_{\text{Tol}}^{-1} \text{Phenol}^{-1} - 0.363\text{Bot}_{\text{Tol}}^{-1} \text{P}_1 - 0.3078\text{Bot}_{\text{Tol}}^{-2} + 0.2738\text{Bot}_{\text{Ph}} - 0.01495\text{Bot}_{\text{Ph}}^{-1} - 0.002238\text{Bot}_{\text{Ph}}^{-2} + 0.4841\text{Bot}_{\text{MCH}}^{-1} \text{Bot}_{\text{Tol}} - 0.2071\text{Bot}_{\text{MCH}}^2 + 0.289\text{Bot}_T \text{Bot}_{\text{Ph}} - 0.1215log(\text{Bot}_{\text{MCH}}) + 0.1959log(\text{Bot}_{\text{Ph}}) + 0.0003221sin(\text{P}_1) - 0.02676sin(\text{P} _{22}) + 0.01961cos(\text{Top}_{\text{Tol}}) + 0.01706sqrt(\text{Bot}_{\text{Ph}})$\\~\\
  $\dot{Q_\text{cond}} = -0.2116\text{V}_{\text{Reb}} \text{P} _{22}^{-1} + 0.01718\text{Bot}_{\text{Tol}} \text{Phenol} + 0.6587\text{Bot}_{\text{Tol}} \text{P} _{22}^{-1} + 1.411\text{Bot}_{\text{MCH}} \text{Bot}_{\text{Tol}}^{-1} + 0.2869\text{Bot}_{\text{MCH}}^2 - 0.1059\text{Bot}_T \text{Bot}_{\text{Tol}}^{-1} - 0.1911\text{Top}_{\text{Tol}} \text{Bot}_{\text{MCH}} - 0.04155\text{Top}_{\text{Ph}}^{-1} \text{Phenol}^{-1} + 0.02561\text{Top}_{\text{Ph}} \text{Bot}_{\text{MCH}}^{-1} - 0.004539\text{Top}_{\text{Ph}}^{-2} + 0.3218\text{Top}_{\text{MCH}}^{-1}Q_\text{cond} + 0.1582\text{Top}_F^{-1} \text{Bot}_{\text{Tol}} + 0.005565sin(\text{Top}_{\text{Tol}}) + 0.006367sin(\text{Bot}_{\text{MCH}})$\\~\\
  $\dot{\text{V}_{\text{Reb}}} = 0.1074\text{Bot}_{\text{Tol}}^{-1} \text{Phenol}^{-1} - 0.1844\text{Bot}_{\text{Tol}} \text{P}_1^{-1} + 1.166\text{Bot}_{\text{Tol}}^{-1} \text{V}_{\text{Reb}} + 0.007661\text{Bot}_{\text{Tol}}^{-2} - 0.1469\text{Bot}_{\text{MCH}}^{-1} \text{Phenol} - 0.5548\text{Bot}_{\text{MCH}}^{-1} \text{V}_{\text{Reb}}^{-1} + 0.09073\text{Bot}_{\text{MCH}} \text{Bot}_{\text{Tol}} - 2.084\text{Bot}_{\text{MCH}} \text{Bot}_{\text{Tol}}^{-1} + 0.3074\text{Bot}_{\text{MCH}}^2 + 0.09942\text{Top}_{\text{Tol}} \text{Bot}_{\text{Tol}} + 0.4745\text{Top}_{\text{Tol}}^{-1} \text{Bot}_{\text{Tol}}^{-1} + 0.562\text{Top}_{\text{Ph}} \text{Top}_{\text{Tol}}^{-1} + 0.0923\text{Top}_{\text{Ph}}^{-2} + 0.1405\text{Top}_T^{-1} \text{V}_{\text{Reb}} - 0.003422sin(\text{Bot}_{\text{Ph}}) + 0.203sin(\text{P} _{22}) + 0.003091cos(\text{Top}_{\text{MCH}}) + 0.004538cos(\text{Bot}_{\text{MCH}}) - 0.004878cos(\text{Bot}_{\text{Ph}}) + 0.008143cos(\text{Bot}_{\text{Tol}})$\\~\\
  $\dot{\text{P}_1} = -0.01488\text{P}_1 \text{Phenol} + 0.02832\text{Bot}_{\text{Tol}}^{-1} \text{Phenol}^{-1} - 0.203\text{Bot}_{\text{Tol}} \text{P} _{22}^{-1} + 0.0414\text{Bot}_{\text{MCH}}^{-1} \text{Bot}_{\text{Tol}} - 0.7068\text{Bot}_{\text{MCH}} \text{Bot}_{\text{Tol}}^{-1} - 0.3569\text{Bot}_{\text{MCH}}^2 + 0.1263\text{Bot}_T \text{Bot}_{\text{Ph}} + 0.04706\text{Top}_{\text{Tol}}^{-1} \text{Bot}_{\text{Tol}} + 0.0852\text{Top}_{\text{Tol}} \text{Bot}_{\text{MCH}} + 0.009267\text{Top}_{\text{Ph}}^{-1} \text{Phenol}^{-1} - 0.08634log(\text{Bot}_{\text{MCH}}) + 0.6433exp(\text{Bot}_{\text{Ph}}) - 0.03791sin(\text{Top}_{\text{Tol}}) - 0.004637sin(\text{Bot}_{\text{MCH}})$\\~\\
  $\dot{\text{P} _{22}} = 0.005539\text{Bot}_{\text{Ph}} + 0.08462\text{Bot}_{\text{Ph}}^2 + 0.04228\text{Bot}_{\text{MCH}}^{-1} \text{P}_1 + 0.1671\text{Bot}_{\text{MCH}}^{-1} \text{Bot}_{\text{Tol}} - .9469\text{Bot}_{\text{MCH}} \text{Bot}_{\text{Tol}}^{-1} - 0.1653\text{Bot}_{\text{MCH}}^2 + 0.2072\text{Bot}_T \text{Bot}_{\text{Ph}} + 0.3954\text{Top}_{\text{Tol}}^{-1} \text{Bot}_{\text{Tol}} + 0.0001409\text{Top}_{\text{Tol}} \text{Bot}_{\text{MCH}} - 0.06709\text{Top}_{\text{Ph}} \text{Phenol} + 0.05785\text{Top}_{\text{Ph}}^{-1} \text{Bot}_{\text{Tol}} + 0.2927log(\text{Bot}_{\text{Ph}}) + 0.1641exp(\text{Bot}_{\text{Ph}}) - 0.03764sin(\text{Top}_{\text{Tol}}) - 0.002023sin(\text{Bot}_{\text{MCH}}) + 0.01122sin(\text{P}_1)$
\section{Common Terms across Systems}
\subsection{Top Stream Flow Rate}
\label{app:ode_common}
\begin{longtable}[c]{|l|l|l|l|l|l|l|l|l|}
\caption{High Regularization - Number of Terms Retained across Systems - $\text{Top}_F$}
\label{tab:terms1}\\
\hline
\cellcolor[HTML]{EFEFEF} & \multicolumn{2}{l|}{\cellcolor[HTML]{EFEFEF}\textbf{Excluding}} & \multicolumn{2}{l|}{\cellcolor[HTML]{EFEFEF}\textbf{MCH400}} & \multicolumn{2}{l|}{\cellcolor[HTML]{EFEFEF}\textbf{T400}} & \multicolumn{2}{l|}{\cellcolor[HTML]{EFEFEF}\textbf{RR6}} \\ \cline{2-9} 
\multirow{-2}{*}{\cellcolor[HTML]{EFEFEF}\textbf{}} & Common & Total & Common & Total & Common & Total & Common & Total \\ \hline
\endfirsthead
\endhead
\textbf{Basic} & 0 & 16 & 6 & 14 & 5 & 14 & 0 & 16 \\ \hline
\textbf{MCH400} & 0 & 14 &  &  & 7 & 19 & 0 & 14 \\ \hline
\textbf{T400} & 0 & 14 &  &  &  &  & 0 & 14 \\ \hline
\textbf{RR6} & 2 & 14 &  &  &  &  &  & \\ \hline
\end{longtable}
\begin{longtable}[c]{|l|l|l|l|l|l|l|l|l|}
\caption{Low Regularization - Number of Terms Retained across Systems}
\label{tab:low_comm1}\\
\hline
\cellcolor[HTML]{EFEFEF} & \multicolumn{2}{l|}{\cellcolor[HTML]{EFEFEF}Excluding} & \multicolumn{2}{l|}{\cellcolor[HTML]{EFEFEF}MCH400} & \multicolumn{2}{l|}{\cellcolor[HTML]{EFEFEF}T400} & \multicolumn{2}{l|}{\cellcolor[HTML]{EFEFEF}RR6} \\ \cline{2-9} 
\multirow{-2}{*}{\cellcolor[HTML]{EFEFEF}} & Common & Total & Common & Total & Common & Total & Common & Total \\ \hline
\endfirsthead
\endhead
Basic & 9 & 37 & 12 & 25 & 10 & 25 & 5 & 25 \\ \hline
MCH400 & 4 & 25 &  &  & 24 & 39 & 13 & 37 \\ \hline
T400 & 4 & 25 &  &  &  &  & 17 & 37 \\ \hline
RR6 & 8 & 25 &  &  &  &  &  &  \\ \hline
\end{longtable}
\begin{longtable}[c]{|l|l|}
\caption{Terms Retained across Systems (High Regularization) - $\text{Top}_{F}$}
\label{tab:terms_1}\\
\hline
\rowcolor[HTML]{EFEFEF} 
\textbf{2 or more} & \textbf{3 or more} \\ \hline
\endfirsthead
\endhead
Bot\_Tol\textasciicircum{}1 Phenol\textasciicircum{}1 & Top\_Tol\textasciicircum{}1 Bot\_MCH\textasciicircum{}1 \\ \hline
Bot\_Tol\textasciicircum{}1 P22\textasciicircum{}-1 & sin(Top\_Tol) \\ \hline
Bot\_MCH\textasciicircum{}1 Bot\_Tol\textasciicircum{}-1 &  \\ \hline
Bot\_MCH\textasciicircum{}2 &  \\ \hline
Top\_Tol\textasciicircum{}-1 Phenol\textasciicircum{}-1 &  \\ \hline
Top\_Tol\textasciicircum{}1 Bot\_MCH\textasciicircum{}1 &  \\ \hline
Top\_F\textasciicircum{}-1 Bot\_Tol\textasciicircum{}1 &  \\ \hline
sin(Top\_Tol) &  \\ \hline
cos(Top\_Tol) &  \\ \hline
P1\textasciicircum{}1 Phenol\textasciicircum{}1 &  \\ \hline
Bot\_MCH\textasciicircum{}1 Phenol\textasciicircum{}1 &  \\ \hline
Top\_Tol\textasciicircum{}-1 Bot\_MCH\textasciicircum{}-1 &  \\ \hline
Top\_Ph\textasciicircum{}-2 &  \\ \hline
sin(P22) &  \\ \hline
\end{longtable}
\subsection{Top Stream Temperature}
\begin{longtable}[c]{|l|l|l|l|l|l|l|l|l|}
\caption{High Regularization - Number of Terms Retained across Systems - $\text{Top}_T$}
\label{tab:terms2}\\
\hline
\cellcolor[HTML]{EFEFEF} & \multicolumn{2}{l|}{\cellcolor[HTML]{EFEFEF}\textbf{Excluding}} & \multicolumn{2}{l|}{\cellcolor[HTML]{EFEFEF}\textbf{MCH400}} & \multicolumn{2}{l|}{\cellcolor[HTML]{EFEFEF}\textbf{T400}} & \multicolumn{2}{l|}{\cellcolor[HTML]{EFEFEF}\textbf{RR6}} \\ \cline{2-9} 
\multirow{-2}{*}{\cellcolor[HTML]{EFEFEF}\textbf{}} & Common & Total & Common & Total & Common & Total & Common & Total \\ \hline
\endfirsthead
\endhead
\textbf{Basic} & 0 & 16 & 6 & 15 & 3 & 15 & 0 & 15 \\ \hline
\textbf{MCH400} & 0 & 15 &  &  & 5 & 18 & 0 & 16 \\ \hline
\textbf{T400} & 0 & 15 &  &  &  &  & 0 & 16 \\ \hline
\textbf{RR6} & 2 & 15 &  &  &  &  &  & \\ \hline
\end{longtable}
\begin{longtable}[c]{|l|l|l|l|l|l|l|l|l|}
\caption{Low Regularization - Number of Terms Retained across Systems}
\label{tab:low_comm2}\\
\hline
\cellcolor[HTML]{EFEFEF} & \multicolumn{2}{l|}{\cellcolor[HTML]{EFEFEF}Excluding} & \multicolumn{2}{l|}{\cellcolor[HTML]{EFEFEF}MCH400} & \multicolumn{2}{l|}{\cellcolor[HTML]{EFEFEF}T400} & \multicolumn{2}{l|}{\cellcolor[HTML]{EFEFEF}RR6} \\ \cline{2-9} 
\multirow{-2}{*}{\cellcolor[HTML]{EFEFEF}} & Common & Total & Common & Total & Common & Total & Common & Total \\ \hline
\endfirsthead
\endhead
Basic & 6 & 36 & 12 & 27 & 14 & 27 & 6 & 27 \\ \hline
MCH400 & 3 & 27 &  &  & 22 & 36 & 11 & 36 \\ \hline
T400 & 4 & 27 &  &  &  &  & 14 & 40 \\ \hline
RR6 & 9 & 27 &  &  &  &  &  &  \\ \hline
\end{longtable}
\begin{longtable}[c]{|l|l|}
\caption{Terms Retained across Systems (High Regularization) - $\text{Top}_{T}$}
\label{tab:terms_2}\\
\hline
\rowcolor[HTML]{EFEFEF} 
\textbf{2 or more} & \textbf{3 or more} \\ \hline
\endfirsthead
\endhead
Bot\_Tol\textasciicircum{}1 Phenol\textasciicircum{}1 & Bot\_Tol\textasciicircum{}1 P22\textasciicircum{}-1 \\ \hline
Bot\_Tol\textasciicircum{}1 P22\textasciicircum{}-1 & Top\_Tol\textasciicircum{}1 Bot\_MCH\textasciicircum{}1 \\ \hline
Bot\_MCH\textasciicircum{}1 Bot\_Tol\textasciicircum{}-1 &  \\ \hline
Top\_Tol\textasciicircum{}-1 Phenol\textasciicircum{}-1 &  \\ \hline
Top\_Tol\textasciicircum{}-1 P1\textasciicircum{}1 &  \\ \hline
Top\_Tol\textasciicircum{}1 Bot\_MCH\textasciicircum{}1 &  \\ \hline
sin(Top\_Tol) &  \\ \hline
P1\textasciicircum{}1 Phenol\textasciicircum{}1 &  \\ \hline
Bot\_MCH\textasciicircum{}1 Phenol\textasciicircum{}1 &  \\ \hline
sin(P22) &  \\ \hline
\end{longtable}
\subsection{Top Stream MCH Flow Rate}
\begin{longtable}[c]{|l|l|l|l|l|l|l|l|l|}
\caption{High Regularization - Number of Terms Retained across Systems - $\text{Top}_{MCH}$}
\label{tab:terms3}\\
\hline
\cellcolor[HTML]{EFEFEF} & \multicolumn{2}{l|}{\cellcolor[HTML]{EFEFEF}\textbf{Excluding}} & \multicolumn{2}{l|}{\cellcolor[HTML]{EFEFEF}\textbf{MCH400}} & \multicolumn{2}{l|}{\cellcolor[HTML]{EFEFEF}\textbf{T400}} & \multicolumn{2}{l|}{\cellcolor[HTML]{EFEFEF}\textbf{RR6}} \\ \cline{2-9} 
\multirow{-2}{*}{\cellcolor[HTML]{EFEFEF}\textbf{}} & Common & Total & Common & Total & Common & Total & Common & Total \\ \hline
\endfirsthead
\endhead
\textbf{Basic} & 0 & 16 & 6 & 14 & 5 & 14 & 0 & 14 \\ \hline
\textbf{MCH400} & 0 & 14 &  &  & 6 & 19 & 0 & 16 \\ \hline
\textbf{T400} & 0 & 14 &  &  &  &  & 0 & 16 \\ \hline
\textbf{RR6} & 1 & 14 &  &  &  &  &  & \\ \hline
\end{longtable}
\begin{longtable}[c]{|l|l|l|l|l|l|l|l|l|}
\caption{Low Regularization - Number of Terms Retained across Systems}
\label{tab:low_comm3}\\
\hline
\cellcolor[HTML]{EFEFEF} & \multicolumn{2}{l|}{\cellcolor[HTML]{EFEFEF}Excluding} & \multicolumn{2}{l|}{\cellcolor[HTML]{EFEFEF}MCH400} & \multicolumn{2}{l|}{\cellcolor[HTML]{EFEFEF}T400} & \multicolumn{2}{l|}{\cellcolor[HTML]{EFEFEF}RR6} \\ \cline{2-9} 
\multirow{-2}{*}{\cellcolor[HTML]{EFEFEF}} & Common & Total & Common & Total & Common & Total & Common & Total \\ \hline
\endfirsthead
\endhead
Basic & 7 & 36 & 14 & 28 & 9 & 28 & 6 & 28 \\ \hline
MCH400 & 3 & 28 &  &  & 23 & 39 & 12 & 36 \\ \hline
T400 & 5 & 28 &  &  &  &  & 14 & 36 \\ \hline
RR6 & 8 & 28 &  &  &  &  &  &  \\ \hline
\end{longtable}
\begin{longtable}[c]{|l|l|}
\caption{Terms Retained across Systems (High Regularization) - $\text{Top}_{MCH}$}
\label{tab:terms_3}\\
\hline
\rowcolor[HTML]{EFEFEF} 
\textbf{2 or more} & \textbf{3 or more} \\ \hline
\endfirsthead
\endhead
Bot\_Tol\textasciicircum{}1 Phenol\textasciicircum{}1 & Top\_Tol\textasciicircum{}1 Bot\_MCH\textasciicircum{}1 \\ \hline
\begin{tabular}[c]{@{}l@{}}Bot\_Tol\textasciicircum{}1\\   P22\textasciicircum{}-1\end{tabular} &  \\ \hline
\begin{tabular}[c]{@{}l@{}}Bot\_MCH\textasciicircum{}1\\   Bot\_Tol\textasciicircum{}-1\end{tabular} &  \\ \hline
Bot\_MCH\textasciicircum{}2 &  \\ \hline
\begin{tabular}[c]{@{}l@{}}Top\_Tol\textasciicircum{}-1\\   Phenol\textasciicircum{}-1\end{tabular} &  \\ \hline
\begin{tabular}[c]{@{}l@{}}Top\_Tol\textasciicircum{}1\\   Bot\_MCH\textasciicircum{}1\end{tabular} &  \\ \hline
\begin{tabular}[c]{@{}l@{}}Top\_Ph\textasciicircum{}-1\\   P1\textasciicircum{}1\end{tabular} &  \\ \hline
\begin{tabular}[c]{@{}l@{}}Top\_F\textasciicircum{}-1\\   Bot\_Tol\textasciicircum{}1\end{tabular} &  \\ \hline
sin(Top\_Tol) &  \\ \hline
cos(Top\_Tol) &  \\ \hline
\begin{tabular}[c]{@{}l@{}}P1\textasciicircum{}1\\   Phenol\textasciicircum{}1\end{tabular} &  \\ \hline
\begin{tabular}[c]{@{}l@{}}Bot\_Tol\textasciicircum{}1\\   P1\textasciicircum{}-1\end{tabular} &  \\ \hline
\begin{tabular}[c]{@{}l@{}}Bot\_MCH\textasciicircum{}1\\   Phenol\textasciicircum{}1\end{tabular} &  \\ \hline
\begin{tabular}[c]{@{}l@{}}Top\_Tol\textasciicircum{}-1\\   Bot\_MCH\textasciicircum{}-1\end{tabular} &  \\ \hline
sin(P22) &  \\ \hline
\end{longtable}
\subsection{Top Stream Phenol Flow Rate}
\begin{longtable}[c]{|l|l|l|l|l|l|l|l|l|}
\caption{High Regularization - Number of Terms Retained across Systems - $\text{Top}_{Ph}$}
\label{tab:terms4}\\
\hline
\cellcolor[HTML]{EFEFEF} & \multicolumn{2}{l|}{\cellcolor[HTML]{EFEFEF}\textbf{Excluding}} & \multicolumn{2}{l|}{\cellcolor[HTML]{EFEFEF}\textbf{MCH400}} & \multicolumn{2}{l|}{\cellcolor[HTML]{EFEFEF}\textbf{T400}} & \multicolumn{2}{l|}{\cellcolor[HTML]{EFEFEF}\textbf{RR6}} \\ \cline{2-9} 
\multirow{-2}{*}{\cellcolor[HTML]{EFEFEF}\textbf{}} & Common & Total & Common & Total & Common & Total & Common & Total \\ \hline
\endfirsthead
\endhead
\textbf{Basic} & 7 & 30 & 2 & 6 & 3 & 6 & 2 & 6 \\ \hline
\textbf{MCH400} & 1 & 6 &  &  & 18 & 34 & 14 & 30 \\ \hline
\textbf{T400} & 1 & 6 &  &  &  &  & 9 & 30 \\ \hline
\textbf{RR6} & 1 & 6 &  &  &  &  & & \\ \hline
\end{longtable}
\begin{longtable}[c]{|l|l|l|l|l|l|l|l|l|}
\caption{Low Regularization - Number of Terms Retained across Systems}
\label{tab:low_comm4}\\
\hline
\cellcolor[HTML]{EFEFEF} & \multicolumn{2}{l|}{\cellcolor[HTML]{EFEFEF}Excluding} & \multicolumn{2}{l|}{\cellcolor[HTML]{EFEFEF}MCH400} & \multicolumn{2}{l|}{\cellcolor[HTML]{EFEFEF}T400} & \multicolumn{2}{l|}{\cellcolor[HTML]{EFEFEF}RR6} \\ \cline{2-9} 
\multirow{-2}{*}{\cellcolor[HTML]{EFEFEF}} & Common & Total & Common & Total & Common & Total & Common & Total \\ \hline
\endfirsthead
\endhead
Basic & 2 & 6 & 4 & 6 & 3 & 6 & 2 & 6 \\ \hline
MCH400 & 2 & 6 &  &  & 24 & 34 & 25 & 43 \\ \hline
T400 & 1 & 6 &  &  &  &  & 16 & 34 \\ \hline
RR6 & 15 & 34 &  &  &  &  &  &  \\ \hline
\end{longtable}
\begin{longtable}[c]{|l|l|}
\caption{Terms Retained across Systems (High Regularization) - $\text{Top}_{Ph}$}
\label{tab:terms_4}\\
\hline
\rowcolor[HTML]{EFEFEF} 
\textbf{2 or more} & \textbf{3 or more} \\ \hline
\endfirsthead
\endhead
Top\_Tol\textasciicircum{}1 Bot\_MCH\textasciicircum{}1 & Top\_Tol\textasciicircum{}2 \\ \hline
Top\_Tol\textasciicircum{}2 & Top\_T\textasciicircum{}1 V\_Reb\textasciicircum{}-1 \\ \hline
Top\_T\textasciicircum{}1 V\_Reb\textasciicircum{}-1 & sin(Top\_Tol) \\ \hline
sin(Top\_Tol) & sin(Bot\_MCH) \\ \hline
P1\textasciicircum{}-1 Phenol\textasciicircum{}-1 & sin(Q\_Cond) \\ \hline
Bot\_Tol\textasciicircum{}-2 & sin(Phenol) \\ \hline
Bot\_MCH\textasciicircum{}1 Phenol\textasciicircum{}-1 & cos(Top\_T) \\ \hline
Bot\_MCH\textasciicircum{}1 Bot\_Tol\textasciicircum{}-1 & cos(Bot\_Ph) \\ \hline
Bot\_MCH\textasciicircum{}2 & cos(Q\_Cond) \\ \hline
Bot\_T\textasciicircum{}1 Bot\_Ph\textasciicircum{}1 & cos(Phenol) \\ \hline
exp(Phenol) &  \\ \hline
sin(Top\_T) &  \\ \hline
sin(Bot\_T) &  \\ \hline
sin(Bot\_MCH) &  \\ \hline
sin(Bot\_Ph) &  \\ \hline
sin(Bot\_Tol) &  \\ \hline
sin(Q\_Cond) &  \\ \hline
sin(V\_Reb) &  \\ \hline
sin(P1) &  \\ \hline
sin(Phenol) &  \\ \hline
cos(Top\_T) &  \\ \hline
cos(Bot\_T) &  \\ \hline
cos(Bot\_Ph) &  \\ \hline
cos(Q\_Cond) &  \\ \hline
cos(V\_Reb) &  \\ \hline
cos(P1) &  \\ \hline
cos(Phenol) &  \\ \hline
\begin{tabular}[c]{@{}l@{}}Top\_Tol\textasciicircum{}1\\   Phenol\textasciicircum{}1\end{tabular} &  \\ \hline
\end{longtable}
\subsection{Top Stream Toluene Flow Rate}
\begin{longtable}[c]{|l|l|l|l|l|l|l|l|l|}
\caption{High Regularization - Number of Terms Retained across Systems - $\text{Top}_{Tol}$}
\label{tab:terms5}\\
\hline
\cellcolor[HTML]{EFEFEF} & \multicolumn{2}{l|}{\cellcolor[HTML]{EFEFEF}\textbf{Excluding}} & \multicolumn{2}{l|}{\cellcolor[HTML]{EFEFEF}\textbf{MCH400}} & \multicolumn{2}{l|}{\cellcolor[HTML]{EFEFEF}\textbf{T400}} & \multicolumn{2}{l|}{\cellcolor[HTML]{EFEFEF}\textbf{RR6}} \\ \cline{2-9} 
\multirow{-2}{*}{\cellcolor[HTML]{EFEFEF}\textbf{}} & Common & Total & Common & Total & Common & Total & Common & Total \\ \hline
\endfirsthead
\endhead
\textbf{Basic} & 0 & 8 & 6 & 9 & 6 & 18 & 3 & 8 \\ \hline
\textbf{MCH400} & 1 & 8 &  &  & 4 & 9 & 2 & 8 \\ \hline
\textbf{T400} & 2 & 8 &  &  &  &  & 1 & 8 \\ \hline
\textbf{RR6} & 4 & 9 &  &  &  &  & & \\ \hline
\end{longtable}
\begin{longtable}[c]{|l|l|l|l|l|l|l|l|l|}
\caption{Low Regularization - Number of Terms Retained across Systems}
\label{tab:low_comm5}\\
\hline
\cellcolor[HTML]{EFEFEF} & \multicolumn{2}{l|}{\cellcolor[HTML]{EFEFEF}Excluding} & \multicolumn{2}{l|}{\cellcolor[HTML]{EFEFEF}MCH400} & \multicolumn{2}{l|}{\cellcolor[HTML]{EFEFEF}T400} & \multicolumn{2}{l|}{\cellcolor[HTML]{EFEFEF}RR6} \\ \cline{2-9} 
\multirow{-2}{*}{\cellcolor[HTML]{EFEFEF}} & Common & Total & Common & Total & Common & Total & Common & Total \\ \hline
\endfirsthead
\endhead
Basic & 1 & 8 & 25 & 41 & 20 & 33 & 4 & 8 \\ \hline
MCH400 & 3 & 8 &  &  & 18 & 33 & 2 & 8 \\ \hline
T400 & 2 & 8 &  &  &  &  & 3 & 8 \\ \hline
RR6 & 12 & 33 &  &  &  &  &  &  \\ \hline
\end{longtable}
\begin{longtable}[c]{|l|l|}
\caption{Terms Retained across Systems (High Regularization) - $\text{Top}_{Tol}$}
\label{tab:terms_5}\\
\hline
\rowcolor[HTML]{EFEFEF} 
\textbf{2 or more} & \textbf{3 or more} \\ \hline
\endfirsthead
\endhead
Phenol\textasciicircum{}-2 & Phenol\textasciicircum{}-2 \\ \hline
Bot\_Tol\textasciicircum{}1 Phenol\textasciicircum{}1 & Bot\_Tol\textasciicircum{}1 Phenol\textasciicircum{}1 \\ \hline
Bot\_MCH\textasciicircum{}1 Phenol\textasciicircum{}-1 & Top\_Tol\textasciicircum{}1 Bot\_MCH\textasciicircum{}1 \\ \hline
Top\_Tol\textasciicircum{}1 Bot\_MCH\textasciicircum{}1 & Top\_Tol\textasciicircum{}1 Bot\_MCH\textasciicircum{}-1 \\ \hline
Top\_Tol\textasciicircum{}1 Bot\_MCH\textasciicircum{}-1 & sin(Top\_Tol) \\ \hline
sin(Top\_Tol) & sin(P22) \\ \hline
sin(P22) & cos(Top\_Tol) \\ \hline
cos(Top\_Tol) &  \\ \hline
\end{longtable}
\subsection{Bottom Stream Temperature}
\begin{longtable}[c]{|l|l|l|l|l|l|l|l|l|}
\caption{High Regularization - Number of Terms Retained across Systems - $\text{Bot}_T$}
\label{tab:terms6}\\
\hline
\cellcolor[HTML]{EFEFEF} & \multicolumn{2}{l|}{\cellcolor[HTML]{EFEFEF}\textbf{Excluding}} & \multicolumn{2}{l|}{\cellcolor[HTML]{EFEFEF}\textbf{MCH400}} & \multicolumn{2}{l|}{\cellcolor[HTML]{EFEFEF}\textbf{T400}} & \multicolumn{2}{l|}{\cellcolor[HTML]{EFEFEF}\textbf{RR6}} \\ \cline{2-9} 
\multirow{-2}{*}{\cellcolor[HTML]{EFEFEF}\textbf{}} & Common & Total & Common & Total & Common & Total & Common & Total \\ \hline
\endfirsthead
\endhead
\textbf{Basic} & 2 & 18 & 2 & 19 & 5 & 19 & 2 & 18 \\ \hline
\textbf{MCH400} & 0 & 18 &  &  & 6 & 23 & 4 & 18 \\ \hline
\textbf{T400} & 0 & 18 &  &  &  &  & 5 & 18 \\ \hline
\textbf{RR6} & 0 & 19 &  &  &  &  & & \\ \hline
\end{longtable}
\begin{longtable}[c]{|l|l|l|l|l|l|l|l|l|}
\caption{Low Regularization - Number of Terms Retained across Systems}
\label{tab:low_comm6}\\
\hline
\cellcolor[HTML]{EFEFEF} & \multicolumn{2}{l|}{\cellcolor[HTML]{EFEFEF}Excluding} & \multicolumn{2}{l|}{\cellcolor[HTML]{EFEFEF}MCH400} & \multicolumn{2}{l|}{\cellcolor[HTML]{EFEFEF}T400} & \multicolumn{2}{l|}{\cellcolor[HTML]{EFEFEF}RR6} \\ \cline{2-9} 
\multirow{-2}{*}{\cellcolor[HTML]{EFEFEF}} & Common & Total & Common & Total & Common & Total & Common & Total \\ \hline
\endfirsthead
\endhead
Basic & 16 & 41 & 25 & 50 & 31 & 50 & 16 & 41 \\ \hline
MCH400 & 13 & 41 &  &  & 37 & 60 & 18 & 41 \\ \hline
T400 & 13 & 41 &  &  &  &  & 21 & 41 \\ \hline
RR6 & 21 & 50 &  &  &  &  &  &  \\ \hline
\end{longtable}
\begin{longtable}[c]{|l|l|}
\caption{Terms Retained across Systems (High Regularization) - $\text{Bot}_{T}$}
\label{tab:terms_6}\\
\hline
\rowcolor[HTML]{EFEFEF} 
\textbf{2 or more} & \textbf{3 or more} \\ \hline
\endfirsthead
\endhead
\rowcolor[HTML]{FFFFFF} 
Bot\_Tol\textasciicircum{}1 Phenol\textasciicircum{}1 & sin(Bot\_T) \\ \hline
\rowcolor[HTML]{FFFFFF} 
Bot\_MCH\textasciicircum{}-1 Phenol\textasciicircum{}-1 & cos(Q\_Cond) \\ \hline
\rowcolor[HTML]{FFFFFF} 
Bot\_MCH\textasciicircum{}2 &  \\ \hline
\rowcolor[HTML]{FFFFFF} 
\begin{tabular}[c]{@{}l@{}}Top\_Tol\textasciicircum{}1\\   Phenol\textasciicircum{}1\end{tabular} &  \\ \hline
\rowcolor[HTML]{FFFFFF} 
\begin{tabular}[c]{@{}l@{}}Top\_Tol\textasciicircum{}-1\\   Bot\_MCH\textasciicircum{}-1\end{tabular} &  \\ \hline
\rowcolor[HTML]{FFFFFF} 
\begin{tabular}[c]{@{}l@{}}Top\_Ph\textasciicircum{}-1\\   Top\_Tol\textasciicircum{}1\end{tabular} &  \\ \hline
\rowcolor[HTML]{FFFFFF} 
cos(Top\_Tol) &  \\ \hline
\rowcolor[HTML]{FFFFFF} 
cos(Bot\_Tol) &  \\ \hline
\rowcolor[HTML]{FFFFFF} 
cos(P22) &  \\ \hline
\rowcolor[HTML]{FFFFFF} 
\begin{tabular}[c]{@{}l@{}}Bot\_MCH\textasciicircum{}-1\\   Bot\_Tol\textasciicircum{}-1\end{tabular} &  \\ \hline
\rowcolor[HTML]{FFFFFF} 
\begin{tabular}[c]{@{}l@{}}Top\_Tol\textasciicircum{}1\\   Bot\_MCH\textasciicircum{}1\end{tabular} & \textbf{} \\ \hline
\rowcolor[HTML]{FFFFFF} 
\begin{tabular}[c]{@{}l@{}}Top\_Tol\textasciicircum{}-1\\   Bot\_MCH\textasciicircum{}1\end{tabular} &  \\ \hline
\rowcolor[HTML]{FFFFFF} 
\begin{tabular}[c]{@{}l@{}}Top\_Ph\textasciicircum{}1\\   Top\_Tol\textasciicircum{}-1\end{tabular} &  \\ \hline
\rowcolor[HTML]{FFFFFF} 
sin(Bot\_T) &  \\ \hline
\rowcolor[HTML]{FFFFFF} 
cos(Bot\_Ph) &  \\ \hline
\rowcolor[HTML]{FFFFFF} 
cos(Q\_Cond) &  \\ \hline
\rowcolor[HTML]{FFFFFF} 
cos(P1) &  \\ \hline
\rowcolor[HTML]{FFFFFF} 
\begin{tabular}[c]{@{}l@{}}Top\_T\textasciicircum{}1\\   V\_Reb\textasciicircum{}-1\end{tabular} &  \\ \hline
\rowcolor[HTML]{FFFFFF} 
sin(Top\_F) &  \\ \hline
\rowcolor[HTML]{FFFFFF} 
cos(Top\_T) &  \\ \hline
\end{longtable}
\subsection{Bottom Stream MCH Flow Rate}
\begin{longtable}[c]{|l|l|l|l|l|l|l|l|l|}
\caption{High Regularization - Number of Terms Retained across Systems - $\text{Bot}_{MCH}$}
\label{tab:terms7}\\
\hline
\cellcolor[HTML]{EFEFEF} & \multicolumn{2}{l|}{\cellcolor[HTML]{EFEFEF}\textbf{Excluding}} & \multicolumn{2}{l|}{\cellcolor[HTML]{EFEFEF}\textbf{MCH400}} & \multicolumn{2}{l|}{\cellcolor[HTML]{EFEFEF}\textbf{T400}} & \multicolumn{2}{l|}{\cellcolor[HTML]{EFEFEF}\textbf{RR6}} \\ \cline{2-9} 
\multirow{-2}{*}{\cellcolor[HTML]{EFEFEF}\textbf{}} & Common & Total & Common & Total & Common & Total & Common & Total \\ \hline
\endfirsthead
\endhead
\textbf{Basic} & 3 & 22 & 6 & 22 & 9 & 22 & 4 & 22 \\ \hline
\textbf{MCH400} & 0 & 22 &  &  & 9 & 22 & 6 & 22 \\ \hline
\textbf{T400} & 1 & 22 &  &  &  &  & 6 & 24 \\ \hline
\textbf{RR6} & 4 & 22 &  &  &  &  & & \\ \hline
\end{longtable}
\begin{longtable}[c]{|l|l|l|l|l|l|l|l|l|}
\caption{Low Regularization - Number of Terms Retained across Systems}
\label{tab:low_comm7}\\
\hline
\cellcolor[HTML]{EFEFEF} & \multicolumn{2}{l|}{\cellcolor[HTML]{EFEFEF}Excluding} & \multicolumn{2}{l|}{\cellcolor[HTML]{EFEFEF}MCH400} & \multicolumn{2}{l|}{\cellcolor[HTML]{EFEFEF}T400} & \multicolumn{2}{l|}{\cellcolor[HTML]{EFEFEF}RR6} \\ \cline{2-9} 
\multirow{-2}{*}{\cellcolor[HTML]{EFEFEF}} & Common & Total & Common & Total & Common & Total & Common & Total \\ \hline
\endfirsthead
\endhead
Basic & 20 & 48 & 30 & 45 & 27 & 45 & 16 & 45 \\ \hline
MCH400 & 14 & 45 &  &  & 34 & 63 & 24 & 48 \\ \hline
T400 & 15 & 45 &  &  &  &  & 26 & 48 \\ \hline
RR6 & 21 & 45 &  &  &  &  &  &  \\ \hline
\end{longtable}
\begin{longtable}[c]{|l|l|}
\caption{Terms Retained across Systems (High Regularization) - $\text{Bot}_{MCH}$}
\label{tab:terms_7}\\
\hline
\rowcolor[HTML]{EFEFEF} 
\textbf{2 or more} & \textbf{3 or more} \\ \hline
\endfirsthead
\endhead
\rowcolor[HTML]{FFFFFF} 
P1\textasciicircum{}-1 Phenol\textasciicircum{}-1 & Bot\_MCH\textasciicircum{}1 Bot\_Tol\textasciicircum{}1 \\ \hline
\rowcolor[HTML]{FFFFFF} 
Bot\_MCH\textasciicircum{}1 Bot\_Tol\textasciicircum{}1 & Bot\_MCH\textasciicircum{}1 Bot\_Tol\textasciicircum{}-1 \\ \hline
\rowcolor[HTML]{FFFFFF} 
Bot\_MCH\textasciicircum{}1 Bot\_Tol\textasciicircum{}-1 & Bot\_MCH\textasciicircum{}2 \\ \hline
Bot\_MCH\textasciicircum{}2 & Top\_Ph\textasciicircum{}1 Top\_Tol\textasciicircum{}-1 \\ \hline
Top\_Ph\textasciicircum{}1 Bot\_MCH\textasciicircum{}1 & Top\_F\textasciicircum{}1 Bot\_Tol\textasciicircum{}-1 \\ \hline
Top\_Ph\textasciicircum{}1 Top\_Tol\textasciicircum{}-1 & sin(Bot\_T) \\ \hline
Top\_MCH\textasciicircum{}1 Bot\_Tol\textasciicircum{}-1 & cos(Bot\_T) \\ \hline
\rowcolor[HTML]{FFFFFF} 
Top\_T\textasciicircum{}1 V\_Reb\textasciicircum{}-1 & cos(Q\_Cond) \\ \hline
\rowcolor[HTML]{FFFFFF} 
\begin{tabular}[c]{@{}l@{}}Top\_F\textasciicircum{}1\\   Bot\_Tol\textasciicircum{}-1\end{tabular} &  \\ \hline
\rowcolor[HTML]{FFFFFF} 
sin(Top\_Tol) &  \\ \hline
\rowcolor[HTML]{FFFFFF} 
sin(P1) &  \\ \hline
\rowcolor[HTML]{FFFFFF} 
cos(Top\_Tol) &  \\ \hline
\rowcolor[HTML]{FFFFFF} 
cos(Bot\_Tol) &  \\ \hline
\rowcolor[HTML]{FFFFFF} 
cos(V\_Reb) &  \\ \hline
\rowcolor[HTML]{FFFFFF} 
\begin{tabular}[c]{@{}l@{}}Bot\_T\textasciicircum{}1\\   Bot\_Ph\textasciicircum{}1\end{tabular} & \textbf{} \\ \hline
\rowcolor[HTML]{FFFFFF} 
sin(Bot\_T) &  \\ \hline
\rowcolor[HTML]{FFFFFF} 
sin(P22) &  \\ \hline
\rowcolor[HTML]{FFFFFF} 
cos(Bot\_T) &  \\ \hline
\rowcolor[HTML]{FFFFFF} 
cos(Bot\_Ph) &  \\ \hline
\rowcolor[HTML]{FFFFFF} 
cos(Q\_Cond) &  \\ \hline
\rowcolor[HTML]{FFFFFF} 
cos(Phenol) &  \\ \hline
\rowcolor[HTML]{FFFFFF} 
cos(Top\_MCH) &  \\ \hline
\rowcolor[HTML]{FFFFFF} 
cos(Bot\_MCH) &  \\ \hline
\rowcolor[HTML]{FFFFFF} 
cos(P22) &  \\ \hline
\end{longtable}
\subsection{Bottom Stream Phenol Flow Rate}
\begin{longtable}[c]{|l|l|l|l|l|l|l|l|l|}
\caption{High Regularization - Number of Terms Retained across Systems - $\text{Bot}_{Ph}$}
\label{tab:terms8}\\
\hline
\cellcolor[HTML]{EFEFEF} & \multicolumn{2}{l|}{\cellcolor[HTML]{EFEFEF}\textbf{Excluding}} & \multicolumn{2}{l|}{\cellcolor[HTML]{EFEFEF}\textbf{MCH400}} & \multicolumn{2}{l|}{\cellcolor[HTML]{EFEFEF}\textbf{T400}} & \multicolumn{2}{l|}{\cellcolor[HTML]{EFEFEF}\textbf{RR6}} \\ \cline{2-9} 
\multirow{-2}{*}{\cellcolor[HTML]{EFEFEF}\textbf{}} & Common & Total & Common & Total & Common & Total & Common & Total \\ \hline
\endfirsthead
\endhead
\textbf{Basic} & 2 & 23 & 5 & 23 & 12 & 23 & 3 & 23 \\ \hline
\textbf{MCH400} & 0 & 23 &  &  & 18 & 23 & 4 & 23 \\ \hline
\textbf{T400} & 1 & 23 &  &  &  &  & 6 & 23 \\ \hline
\textbf{RR6} & 3 & 23 &  &  &  &  & & \\ \hline
\end{longtable}
\begin{longtable}[c]{|l|l|l|l|l|l|l|l|l|}
\caption{Low Regularization - Number of Terms Retained across Systems}
\label{tab:low_comm8}\\
\hline
\cellcolor[HTML]{EFEFEF} & \multicolumn{2}{l|}{\cellcolor[HTML]{EFEFEF}Excluding} & \multicolumn{2}{l|}{\cellcolor[HTML]{EFEFEF}MCH400} & \multicolumn{2}{l|}{\cellcolor[HTML]{EFEFEF}T400} & \multicolumn{2}{l|}{\cellcolor[HTML]{EFEFEF}RR6} \\ \cline{2-9} 
\multirow{-2}{*}{\cellcolor[HTML]{EFEFEF}} & Common & Total & Common & Total & Common & Total & Common & Total \\ \hline
\endfirsthead
\endhead
Basic & 18 & 48 & 32 & 57 & 37 & 57 & 24 & 48 \\ \hline
MCH400 & 17 & 48 &  &  & 47 & 78 & 23 & 48 \\ \hline
T400 & 19 & 48 &  &  &  &  & 22 & 48 \\ \hline
RR6 & 28 & 57 &  &  &  &  &  &  \\ \hline
\end{longtable}
\begin{longtable}[c]{|l|l|}
\caption{Terms Retained across Systems (High Regularization) - $\text{Bot}_{Ph}$}
\label{tab:terms_8}\\
\hline
\rowcolor[HTML]{EFEFEF} 
\textbf{2 or more} & \textbf{3 or more} \\ \hline
\endfirsthead
\endhead
\rowcolor[HTML]{FFFFFF} 
Bot\_Tol\textasciicircum{}1 Phenol\textasciicircum{}1 & Bot\_MCH\textasciicircum{}2 \\ \hline
\rowcolor[HTML]{FFFFFF} 
Bot\_Tol\textasciicircum{}-1 P1\textasciicircum{}1 & Top\_T\textasciicircum{}-1 V\_Reb\textasciicircum{}1 \\ \hline
\rowcolor[HTML]{FFFFFF} 
Bot\_Tol\textasciicircum{}1 P1\textasciicircum{}-1 & Top\_F\textasciicircum{}1 Bot\_Tol\textasciicircum{}-1 \\ \hline
Bot\_Tol\textasciicircum{}2 & sin(Q\_Cond) \\ \hline
Bot\_MCH\textasciicircum{}-1 Phenol\textasciicircum{}-1 & Top\_Ph\textasciicircum{}1 Top\_Tol\textasciicircum{}-1 \\ \hline
Bot\_MCH\textasciicircum{}1 Bot\_Tol\textasciicircum{}1 & sin(Bot\_T) \\ \hline
Bot\_MCH\textasciicircum{}2 &  \\ \hline
\rowcolor[HTML]{FFFFFF} 
\begin{tabular}[c]{@{}l@{}}Top\_Tol\textasciicircum{}-1\\   Bot\_Tol\textasciicircum{}1\end{tabular} &  \\ \hline
\rowcolor[HTML]{FFFFFF} 
\begin{tabular}[c]{@{}l@{}}Top\_Tol\textasciicircum{}-1\\   Bot\_MCH\textasciicircum{}-1\end{tabular} &  \\ \hline
\rowcolor[HTML]{FFFFFF} 
\begin{tabular}[c]{@{}l@{}}Top\_T\textasciicircum{}-1\\   V\_Reb\textasciicircum{}1\end{tabular} &  \\ \hline
\rowcolor[HTML]{FFFFFF} 
\begin{tabular}[c]{@{}l@{}}Top\_F\textasciicircum{}1\\   Bot\_Tol\textasciicircum{}-1\end{tabular} &  \\ \hline
\rowcolor[HTML]{FFFFFF} 
sin(Top\_Tol) &  \\ \hline
\rowcolor[HTML]{FFFFFF} 
sin(Bot\_MCH) &  \\ \hline
\rowcolor[HTML]{FFFFFF} 
sin(Q\_Cond) &  \\ \hline
\rowcolor[HTML]{FFFFFF} 
cos(Top\_Tol) & \textbf{} \\ \hline
\rowcolor[HTML]{FFFFFF} 
cos(Bot\_Tol) &  \\ \hline
\rowcolor[HTML]{FFFFFF} 
\begin{tabular}[c]{@{}l@{}}Q\_Cond\textasciicircum{}1\\   P22\textasciicircum{}-1\end{tabular} &  \\ \hline
\rowcolor[HTML]{FFFFFF} 
\begin{tabular}[c]{@{}l@{}}Bot\_Tol\textasciicircum{}-1\\   Phenol\textasciicircum{}-1\end{tabular} &  \\ \hline
\rowcolor[HTML]{FFFFFF} 
Bot\_MCH\textasciicircum{}-2 &  \\ \hline
\rowcolor[HTML]{FFFFFF} 
\begin{tabular}[c]{@{}l@{}}Bot\_T\textasciicircum{}1\\   Bot\_Ph\textasciicircum{}1\end{tabular} &  \\ \hline
\rowcolor[HTML]{FFFFFF} 
\begin{tabular}[c]{@{}l@{}}Top\_Tol\textasciicircum{}1\\   Bot\_Tol\textasciicircum{}-1\end{tabular} &  \\ \hline
\rowcolor[HTML]{FFFFFF} 
\begin{tabular}[c]{@{}l@{}}Top\_Tol\textasciicircum{}-1\\   Bot\_MCH\textasciicircum{}1\end{tabular} &  \\ \hline
\rowcolor[HTML]{FFFFFF} 
Top\_Tol\textasciicircum{}-2 &  \\ \hline
\begin{tabular}[c]{@{}l@{}}Top\_Ph\textasciicircum{}1\\   Top\_Tol\textasciicircum{}-1\end{tabular} &  \\ \hline
\begin{tabular}[c]{@{}l@{}}Top\_MCH\textasciicircum{}1\\   Bot\_Tol\textasciicircum{}-1\end{tabular} &  \\ \hline
sin(Bot\_T) &  \\ \hline
sin(Bot\_Tol) &  \\ \hline
sin(P1) &  \\ \hline
sin(P22) &  \\ \hline
cos(Top\_MCH) &  \\ \hline
cos(Bot\_T) &  \\ \hline
cos(Bot\_Ph) &  \\ \hline
\begin{tabular}[c]{@{}l@{}}Q\_Cond\textasciicircum{}-1\\   P22\textasciicircum{}1\end{tabular} &  \\ \hline
sin(Top\_MCH) &  \\ \hline
sin(V\_Reb) &  \\ \hline
\rowcolor[HTML]{FFFFFF} 
cos(Phenol) &  \\ \hline
\end{longtable}
\subsection{Bottom Stream Toluene Flow Rate}
\begin{longtable}[c]{|l|l|l|l|l|l|l|l|l|}
\caption{High Regularization - Number of Terms Retained across Systems - $\text{Bot}_{Tol}$}
\label{tab:terms9}\\
\hline
\cellcolor[HTML]{EFEFEF} & \multicolumn{2}{l|}{\cellcolor[HTML]{EFEFEF}\textbf{Excluding}} & \multicolumn{2}{l|}{\cellcolor[HTML]{EFEFEF}\textbf{MCH400}} & \multicolumn{2}{l|}{\cellcolor[HTML]{EFEFEF}\textbf{T400}} & \multicolumn{2}{l|}{\cellcolor[HTML]{EFEFEF}\textbf{RR6}} \\ \cline{2-9} 
\multirow{-2}{*}{\cellcolor[HTML]{EFEFEF}\textbf{}} & Common & Total & Common & Total & Common & Total & Common & Total \\ \hline
\endfirsthead
\endhead
\textbf{Basic} & 1 & 14 & 4 & 21 & 13 & 23 & 0 & 14 \\ \hline
\textbf{MCH400} & 0 & 14 &  &  & 15 & 21 & 2 & 14 \\ \hline
\textbf{T400} & 0 & 14 &  &  &  &  & 2 & 14 \\ \hline
\textbf{RR6} & 4 & 21 &  &  &  &  & & \\ \hline
\end{longtable}
\begin{longtable}[c]{|l|l|l|l|l|l|l|l|l|}
\caption{Low Regularization - Number of Terms Retained across Systems}
\label{tab:low_comm9}\\
\hline
\cellcolor[HTML]{EFEFEF} & \multicolumn{2}{l|}{\cellcolor[HTML]{EFEFEF}Excluding} & \multicolumn{2}{l|}{\cellcolor[HTML]{EFEFEF}MCH400} & \multicolumn{2}{l|}{\cellcolor[HTML]{EFEFEF}T400} & \multicolumn{2}{l|}{\cellcolor[HTML]{EFEFEF}RR6} \\ \cline{2-9} 
\multirow{-2}{*}{\cellcolor[HTML]{EFEFEF}} & Common & Total & Common & Total & Common & Total & Common & Total \\ \hline
\endfirsthead
\endhead
Basic & 10 & 30 & 29 & 50 & 32 & 50 & 10 & 30 \\ \hline
MCH400 & 7 & 30 &  &  & 45 & 63 & 12 & 30 \\ \hline
T400 & 8 & 30 &  &  &  &  & 11 & 30 \\ \hline
RR6 & 24 & 50 &  &  &  &  &  &  \\ \hline
\end{longtable}
\begin{longtable}[c]{|l|l|}
\caption{Terms Retained across Systems (High Regularization) - $\text{Bot}_{Tol}$}
\label{tab:terms_9}\\
\hline
\rowcolor[HTML]{EFEFEF} 
\textbf{2 or more} & \textbf{3 or more} \\ \hline
\endfirsthead
\endhead
\rowcolor[HTML]{FFFFFF} 
Bot\_Tol\textasciicircum{}-1 Phenol\textasciicircum{}-1 & Bot\_Tol\textasciicircum{}-1 Phenol\textasciicircum{}-1 \\ \hline
\rowcolor[HTML]{FFFFFF} 
Bot\_Tol\textasciicircum{}2 & Bot\_T\textasciicircum{}1 Bot\_Ph\textasciicircum{}1 \\ \hline
\rowcolor[HTML]{FFFFFF} 
Bot\_MCH\textasciicircum{}1 Bot\_Tol\textasciicircum{}-1 & sin(Top\_Tol) \\ \hline
Bot\_MCH\textasciicircum{}2 & sin(Q\_Cond) \\ \hline
Bot\_T\textasciicircum{}1 Bot\_Ph\textasciicircum{}1 & Top\_Ph\textasciicircum{}1 Top\_Tol\textasciicircum{}-1 \\ \hline
log(Bot\_MCH) &  \\ \hline
sin(Top\_T) &  \\ \hline
\rowcolor[HTML]{FFFFFF} 
sin(Top\_MCH) &  \\ \hline
\rowcolor[HTML]{FFFFFF} 
sin(Top\_Tol) &  \\ \hline
\rowcolor[HTML]{FFFFFF} 
sin(Q\_Cond) &  \\ \hline
\rowcolor[HTML]{FFFFFF} 
sin(V\_Reb) &  \\ \hline
\rowcolor[HTML]{FFFFFF} 
sin(P22) &  \\ \hline
\rowcolor[HTML]{FFFFFF} 
cos(Bot\_MCH) &  \\ \hline
\rowcolor[HTML]{FFFFFF} 
\begin{tabular}[c]{@{}l@{}}Q\_Cond\textasciicircum{}1\\   P22\textasciicircum{}-1\end{tabular} &  \\ \hline
\rowcolor[HTML]{FFFFFF} 
\begin{tabular}[c]{@{}l@{}}Bot\_Tol\textasciicircum{}1\\   Phenol\textasciicircum{}1\end{tabular} & \textbf{} \\ \hline
\rowcolor[HTML]{FFFFFF} 
\begin{tabular}[c]{@{}l@{}}Bot\_MCH\textasciicircum{}1\\   Phenol\textasciicircum{}1\end{tabular} &  \\ \hline
\rowcolor[HTML]{FFFFFF} 
\begin{tabular}[c]{@{}l@{}}Bot\_MCH\textasciicircum{}-1\\   Phenol\textasciicircum{}1\end{tabular} &  \\ \hline
\rowcolor[HTML]{FFFFFF} 
\begin{tabular}[c]{@{}l@{}}Top\_Tol\textasciicircum{}-1\\   Bot\_MCH\textasciicircum{}1\end{tabular} &  \\ \hline
\rowcolor[HTML]{FFFFFF} 
\begin{tabular}[c]{@{}l@{}}Top\_Ph\textasciicircum{}1\\   Top\_Tol\textasciicircum{}-1\end{tabular} &  \\ \hline
\rowcolor[HTML]{FFFFFF} 
\begin{tabular}[c]{@{}l@{}}Top\_T\textasciicircum{}1\\   V\_Reb\textasciicircum{}-1\end{tabular} &  \\ \hline
\rowcolor[HTML]{FFFFFF} 
sin(Bot\_T) &  \\ \hline
\rowcolor[HTML]{FFFFFF} 
sin(Bot\_Ph) &  \\ \hline
\rowcolor[HTML]{FFFFFF} 
sin(Bot\_Tol) &  \\ \hline
sin(P1) &  \\ \hline
cos(Bot\_Ph) &  \\ \hline
Top\_Ph\textasciicircum{}2 &  \\ \hline
\end{longtable}
\subsection{Condenser Duty}
\begin{longtable}[c]{|l|l|l|l|l|l|l|l|l|}
\caption{High Regularization - Number of Terms Retained across Systems - $\text{Q}_{\text{cond}}$}
\label{tab:terms10}\\
\hline
\cellcolor[HTML]{EFEFEF} & \multicolumn{2}{l|}{\cellcolor[HTML]{EFEFEF}\textbf{Excluding}} & \multicolumn{2}{l|}{\cellcolor[HTML]{EFEFEF}\textbf{MCH400}} & \multicolumn{2}{l|}{\cellcolor[HTML]{EFEFEF}\textbf{T400}} & \multicolumn{2}{l|}{\cellcolor[HTML]{EFEFEF}\textbf{RR6}} \\ \cline{2-9} 
\multirow{-2}{*}{\cellcolor[HTML]{EFEFEF}\textbf{}} & Common & Total & Common & Total & Common & Total & Common & Total \\ \hline
\endfirsthead
\endhead
\textbf{Basic} & 1 & 15 & 6 & 14 & 4 & 14 & 1 & 14 \\ \hline
\textbf{MCH400} & 0 & 14 &  &  & 5 & 18 & 2 & 15 \\ \hline
\textbf{T400} & 1 & 14 &  &  &  &  & 1 & 15 \\ \hline
\textbf{RR6} & 2 & 14 &  &  &  &  & & \\ \hline
\end{longtable}
\begin{longtable}[c]{|l|l|l|l|l|l|l|l|l|}
\caption{Low Regularization - Number of Terms Retained across Systems}
\label{tab:low_comm10}\\
\hline
\cellcolor[HTML]{EFEFEF} & \multicolumn{2}{l|}{\cellcolor[HTML]{EFEFEF}Excluding} & \multicolumn{2}{l|}{\cellcolor[HTML]{EFEFEF}MCH400} & \multicolumn{2}{l|}{\cellcolor[HTML]{EFEFEF}T400} & \multicolumn{2}{l|}{\cellcolor[HTML]{EFEFEF}RR6} \\ \cline{2-9} 
\multirow{-2}{*}{\cellcolor[HTML]{EFEFEF}} & Common & Total & Common & Total & Common & Total & Common & Total \\ \hline
\endfirsthead
\endhead
Basic & 18 & 50 &  & 36 & 21 & 36 & 16 & 36 \\ \hline
MCH400 & 12 & 36 &  &  & 31 & 50 & 22 & 50 \\ \hline
T400 & 14 & 36 &  &  &  &  & 23 & 50 \\ \hline
RR6 & 19 & 36 & All 4 & 11 & 36 &  &  &  \\ \hline
\end{longtable}
\begin{longtable}[c]{|l|l|}
\caption{Terms Retained across Systems (High Regularization) - $\text{Q}_{\text{Cond}}$}
\label{tab:terms_10}\\
\hline
\rowcolor[HTML]{EFEFEF} 
\textbf{2 or more} & \textbf{3 or more} \\ \hline
\endfirsthead
\endhead
\rowcolor[HTML]{FFFFFF} 
Bot\_Tol\textasciicircum{}1 Phenol\textasciicircum{}1 & Bot\_Tol\textasciicircum{}1 P22\textasciicircum{}-1 \\ \hline
\rowcolor[HTML]{FFFFFF} 
Bot\_Tol\textasciicircum{}1 P22\textasciicircum{}-1 & Top\_Tol\textasciicircum{}1 Bot\_MCH\textasciicircum{}1 \\ \hline
\rowcolor[HTML]{FFFFFF} 
Bot\_MCH\textasciicircum{}1 Bot\_Tol\textasciicircum{}-1 & Top\_MCH\textasciicircum{}-1 Q\_Cond\textasciicircum{}1 \\ \hline
Top\_Tol\textasciicircum{}1 Bot\_MCH\textasciicircum{}1 & P1\textasciicircum{}1 Phenol\textasciicircum{}1 \\ \hline
Top\_Ph\textasciicircum{}-2 &  \\ \hline
\begin{tabular}[c]{@{}l@{}}Top\_MCH\textasciicircum{}-1\\   Q\_Cond\textasciicircum{}1\end{tabular} &  \\ \hline
\begin{tabular}[c]{@{}l@{}}Top\_F\textasciicircum{}-1\\   Bot\_Tol\textasciicircum{}1\end{tabular} &  \\ \hline
\rowcolor[HTML]{FFFFFF} 
sin(Top\_Tol) &  \\ \hline
\rowcolor[HTML]{FFFFFF} 
\begin{tabular}[c]{@{}l@{}}P1\textasciicircum{}1\\   Phenol\textasciicircum{}1\end{tabular} &  \\ \hline
\rowcolor[HTML]{FFFFFF} 
\begin{tabular}[c]{@{}l@{}}Bot\_MCH\textasciicircum{}1\\   Phenol\textasciicircum{}1\end{tabular} &  \\ \hline
\rowcolor[HTML]{FFFFFF} 
sin(P22) &  \\ \hline
\end{longtable}
\subsection{Reboiler Vapor Flow Rate}
\begin{longtable}[c]{|l|l|l|l|l|l|l|l|l|}
\caption{High Regularization - Number of Terms Retained across Systems - $\text{Vap}_{Reb}$}
\label{tab:terms11}\\
\hline
\cellcolor[HTML]{EFEFEF} & \multicolumn{2}{l|}{\cellcolor[HTML]{EFEFEF}\textbf{Excluding}} & \multicolumn{2}{l|}{\cellcolor[HTML]{EFEFEF}\textbf{MCH400}} & \multicolumn{2}{l|}{\cellcolor[HTML]{EFEFEF}\textbf{T400}} & \multicolumn{2}{l|}{\cellcolor[HTML]{EFEFEF}\textbf{RR6}} \\ \cline{2-9} 
\multirow{-2}{*}{\cellcolor[HTML]{EFEFEF}\textbf{}} & Common & Total & Common & Total & Common & Total & Common & Total \\ \hline
\endfirsthead
\endhead
\textbf{Basic} & 2 & 20 & 8 & 20 & 11 & 20 & 5 & 20 \\ \hline
\textbf{MCH400} & 4 & 20 &  &  & 6 & 20 & 4 & 20 \\ \hline
\textbf{T400} & 3 & 20 &  &  &  &  & 7 & 21 \\ \hline
\textbf{RR6} & 3 & 20 &  &  &  &  & & \\ \hline
\end{longtable}
\begin{longtable}[c]{|l|l|l|l|l|l|l|l|l|}
\caption{Low Regularization - Number of Terms Retained across Systems}
\label{tab:low_comm11}\\
\hline
\cellcolor[HTML]{EFEFEF} & \multicolumn{2}{l|}{\cellcolor[HTML]{EFEFEF}Excluding} & \multicolumn{2}{l|}{\cellcolor[HTML]{EFEFEF}MCH400} & \multicolumn{2}{l|}{\cellcolor[HTML]{EFEFEF}T400} & \multicolumn{2}{l|}{\cellcolor[HTML]{EFEFEF}RR6} \\ \cline{2-9} 
\multirow{-2}{*}{\cellcolor[HTML]{EFEFEF}} & Common & Total & Common & Total & Common & Total & Common & Total \\ \hline
\endfirsthead
\endhead
Basic & 18 & 51 & 24 & 50 & 28 & 50 & 17 & 50 \\ \hline
MCH400 & 13 & 50 &  &  & 38 & 64 & 23 & 51 \\ \hline
T400 & 11 & 50 &  &  &  &  & 23 & 51 \\ \hline
RR6 & 18 & 50 &  &  &  &  &  &  \\ \hline
\end{longtable}
\begin{longtable}[c]{|l|l|}
\caption{Terms Retained across Systems (High Regularization) - $\text{Vap}_{\text{Reb}}$}
\label{tab:terms_11}\\
\hline
\rowcolor[HTML]{EFEFEF} 
\textbf{2 or more} & \textbf{3 or more} \\ \hline
\endfirsthead
\endhead
\rowcolor[HTML]{FFFFFF} 
Bot\_Tol\textasciicircum{}-1 Phenol\textasciicircum{}-1 & Top\_Ph\textasciicircum{}1 Top\_Tol\textasciicircum{}-1 \\ \hline
\rowcolor[HTML]{FFFFFF} 
Bot\_Tol\textasciicircum{}1 P1\textasciicircum{}-1 & Top\_T\textasciicircum{}-1 V\_Reb\textasciicircum{}1 \\ \hline
\rowcolor[HTML]{FFFFFF} 
Bot\_Tol\textasciicircum{}-1 V\_Reb\textasciicircum{}1 & sin(P22) \\ \hline
Bot\_MCH\textasciicircum{}-1 Phenol\textasciicircum{}1 & cos(Top\_MCH) \\ \hline
Bot\_MCH\textasciicircum{}1 Bot\_Tol\textasciicircum{}1 & cos(Bot\_MCH) \\ \hline
Bot\_MCH\textasciicircum{}1 Bot\_Tol\textasciicircum{}-1 & cos(Bot\_Ph) \\ \hline
Bot\_MCH\textasciicircum{}2 &  \\ \hline
\rowcolor[HTML]{FFFFFF} 
\begin{tabular}[c]{@{}l@{}}Top\_Tol\textasciicircum{}-1\\   Bot\_Tol\textasciicircum{}-1\end{tabular} &  \\ \hline
\rowcolor[HTML]{FFFFFF} 
\begin{tabular}[c]{@{}l@{}}Top\_Ph\textasciicircum{}1\\   Top\_Tol\textasciicircum{}-1\end{tabular} &  \\ \hline
\rowcolor[HTML]{FFFFFF} 
\begin{tabular}[c]{@{}l@{}}Top\_T\textasciicircum{}-1\\   V\_Reb\textasciicircum{}1\end{tabular} &  \\ \hline
\rowcolor[HTML]{FFFFFF} 
sin(Bot\_Ph) &  \\ \hline
sin(P22) &  \\ \hline
cos(Top\_MCH) &  \\ \hline
cos(Bot\_MCH) &  \\ \hline
cos(Bot\_Ph) &  \\ \hline
cos(Bot\_Tol) &  \\ \hline
\begin{tabular}[c]{@{}l@{}}Top\_Tol\textasciicircum{}1\\   Bot\_MCH\textasciicircum{}1\end{tabular} &  \\ \hline
\begin{tabular}[c]{@{}l@{}}Top\_Tol\textasciicircum{}-1\\   Bot\_MCH\textasciicircum{}-1\end{tabular} &  \\ \hline
sin(Bot\_T) &  \\ \hline
sin(Q\_Cond) &  \\ \hline
sin(Bot\_MCH) &  \\ \hline
cos(V\_Reb) &  \\ \hline
cos(Phenol) &  \\ \hline
\end{longtable}
\subsection{Pressure Stage 1 (Condenser)}
\begin{longtable}[c]{|l|l|l|l|l|l|l|l|l|}
\caption{High Regularization - Number of Terms Retained across Systems - $\text{P}_{1}$}
\label{tab:terms12}\\
\hline
\cellcolor[HTML]{EFEFEF} & \multicolumn{2}{l|}{\cellcolor[HTML]{EFEFEF}\textbf{Excluding}} & \multicolumn{2}{l|}{\cellcolor[HTML]{EFEFEF}\textbf{MCH400}} & \multicolumn{2}{l|}{\cellcolor[HTML]{EFEFEF}\textbf{T400}} & \multicolumn{2}{l|}{\cellcolor[HTML]{EFEFEF}\textbf{RR6}} \\ \cline{2-9} 
\multirow{-2}{*}{\cellcolor[HTML]{EFEFEF}\textbf{}} & Common & Total & Common & Total & Common & Total & Common & Total \\ \hline
\endfirsthead
\endhead
\textbf{Basic} & 0 & 19 & 8 & 17 & 7 & 17 & 2 & 15 \\ \hline
\textbf{MCH400} & 1 & 17 &  &  & 7 & 18 & 3 & 18 \\ \hline
\textbf{T400} & 0 & 17 &  &  &  &  & 0 & 21 \\ \hline
\textbf{RR6} & 4 & 17 &  &  &  &  & & \\ \hline
\end{longtable}
\begin{longtable}[c]{|l|l|l|l|l|l|l|l|l|}
\caption{Low Regularization - Number of Terms Retained across Systems}
\label{tab:low_comm12}\\
\hline
\cellcolor[HTML]{EFEFEF} & \multicolumn{2}{l|}{\cellcolor[HTML]{EFEFEF}Excluding} & \multicolumn{2}{l|}{\cellcolor[HTML]{EFEFEF}MCH400} & \multicolumn{2}{l|}{\cellcolor[HTML]{EFEFEF}T400} & \multicolumn{2}{l|}{\cellcolor[HTML]{EFEFEF}RR6} \\ \cline{2-9} 
\multirow{-2}{*}{\cellcolor[HTML]{EFEFEF}} & Common & Total & Common & Total & Common & Total & Common & Total \\ \hline
\endfirsthead
\endhead
Basic & 9 & 32 & 14 & 30 & 15 & 30 & 13 & 30 \\ \hline
MCH400 & 7 & 30 &  &  & 22 & 32 & 10 & 32 \\ \hline
T400 & 5 & 30 &  &  &  &  & 18 & 43 \\ \hline
RR6 & 10 & 30 &  &  &  &  &  &  \\ \hline
\end{longtable}
\begin{longtable}[c]{|l|l|}
\caption{Terms Retained across Systems (High Regularization) - $\text{P}_{1}$}
\label{tab:terms_12}\\
\hline
\rowcolor[HTML]{EFEFEF} 
\textbf{\begin{tabular}[c]{@{}l@{}}2 or\\   more\end{tabular}} & \textbf{3 or more} \\ \hline
\endfirsthead
\endhead
\rowcolor[HTML]{FFFFFF} 
Bot\_Tol\textasciicircum{}-1 Phenol\textasciicircum{}-1 & Bot\_Tol\textasciicircum{}1 P22\textasciicircum{}-1 \\ \hline
\rowcolor[HTML]{FFFFFF} 
Bot\_Tol\textasciicircum{}1 P22\textasciicircum{}-1 & Bot\_MCH\textasciicircum{}1 Bot\_Tol\textasciicircum{}-1 \\ \hline
\rowcolor[HTML]{FFFFFF} 
Bot\_MCH\textasciicircum{}1 Bot\_Tol\textasciicircum{}-1 & Top\_Tol\textasciicircum{}-1 Bot\_Tol\textasciicircum{}1 \\ \hline
\rowcolor[HTML]{FFFFFF} 
Bot\_MCH\textasciicircum{}2 & Top\_Tol\textasciicircum{}1 Bot\_MCH\textasciicircum{}1 \\ \hline
\rowcolor[HTML]{FFFFFF} 
Top\_Tol\textasciicircum{}-1 Bot\_Tol\textasciicircum{}1 & sin(Bot\_MCH) \\ \hline
\rowcolor[HTML]{FFFFFF} 
Top\_Tol\textasciicircum{}1 Bot\_MCH\textasciicircum{}1 &  \\ \hline
\rowcolor[HTML]{FFFFFF} 
Top\_Ph\textasciicircum{}-1 P1\textasciicircum{}1 &  \\ \hline
\rowcolor[HTML]{FFFFFF} 
Top\_Ph\textasciicircum{}-1 Top\_Tol\textasciicircum{}-1 &  \\ \hline
\rowcolor[HTML]{FFFFFF} 
Top\_Ph\textasciicircum{}-2 &  \\ \hline
\rowcolor[HTML]{FFFFFF} 
log(Bot\_MCH) &  \\ \hline
\rowcolor[HTML]{FFFFFF} 
sin(Top\_Tol) &  \\ \hline
\rowcolor[HTML]{FFFFFF} 
sin(Bot\_MCH) &  \\ \hline
\rowcolor[HTML]{FFFFFF} 
Bot\_MCH\textasciicircum{}-1 Phenol\textasciicircum{}1 &  \\ \hline
\rowcolor[HTML]{FFFFFF} 
Bot\_T\textasciicircum{}-1 Bot\_Ph\textasciicircum{}-1 &  \\ \hline
\rowcolor[HTML]{FFFFFF} 
Top\_Ph\textasciicircum{}1 Top\_Tol\textasciicircum{}-1 &  \\ \hline
\rowcolor[HTML]{FFFFFF} 
sin(P22) &  \\ \hline
\end{longtable}
\subsection{Pressure Stage 22 (Reboiler)}
\begin{longtable}[c]{|l|l|l|l|l|l|l|l|l|}
\caption{High Regularization - Number of Terms Retained across Systems - $\text{P}_{22}$}
\label{tab:terms13}\\
\hline
\cellcolor[HTML]{EFEFEF} & \multicolumn{2}{l|}{\cellcolor[HTML]{EFEFEF}\textbf{Excluding}} & \multicolumn{2}{l|}{\cellcolor[HTML]{EFEFEF}\textbf{MCH400}} & \multicolumn{2}{l|}{\cellcolor[HTML]{EFEFEF}\textbf{T400}} & \multicolumn{2}{l|}{\cellcolor[HTML]{EFEFEF}\textbf{RR6}} \\ \cline{2-9} 
\multirow{-2}{*}{\cellcolor[HTML]{EFEFEF}\textbf{}} & Common & Total & Common & Total & Common & Total & Common & Total \\ \hline
\endfirsthead
\endhead
\textbf{Basic} & 0 & 15 & 3 & 15 & 3 & 15 & 2 & 15 \\ \hline
\textbf{MCH400} & 0 & 15 &  &  & 7 & 18 & 3 & 18 \\ \hline
\textbf{T400} & 1 & 18 &  &  &  &  & 0 & 21 \\ \hline
\textbf{RR6} & 2 & 15 &  &  &  &  & & \\ \hline
\end{longtable}
\begin{longtable}[c]{|l|l|l|l|l|l|l|l|l|}
\caption{Low Regularization - Number of Terms Retained across Systems}
\label{tab:low_comm13}\\
\hline
\cellcolor[HTML]{EFEFEF} & \multicolumn{2}{l|}{\cellcolor[HTML]{EFEFEF}Excluding} & \multicolumn{2}{l|}{\cellcolor[HTML]{EFEFEF}MCH400} & \multicolumn{2}{l|}{\cellcolor[HTML]{EFEFEF}T400} & \multicolumn{2}{l|}{\cellcolor[HTML]{EFEFEF}RR6} \\ \cline{2-9} 
\multirow{-2}{*}{\cellcolor[HTML]{EFEFEF}} & Common & Total & Common & Total & Common & Total & Common & Total \\ \hline
\endfirsthead
\endhead
Basic & 13 & 38 & 23 & 40 & 23 & 40 & 16 & 38 \\ \hline
MCH400 & 10 & 38 &  &  & 27 & 51 & 19 & 38 \\ \hline
T400 & 13 & 38 &  &  &  &  & 16 & 38 \\ \hline
RR6 & 16 & 40 &  &  &  &  &  &  \\ \hline
\end{longtable}
\begin{longtable}[c]{|l|l|}
\caption{Terms Retained across Systems (High Regularization) - $\text{P}_{22}$}
\label{tab:terms_13}\\
\hline
\rowcolor[HTML]{EFEFEF} 
\textbf{2 or more} & \textbf{3 or more} \\ \hline
\endfirsthead
\endhead
\rowcolor[HTML]{FFFFFF} 
Bot\_MCH\textasciicircum{}1 Bot\_Tol\textasciicircum{}-1 & Bot\_MCH\textasciicircum{}1 Bot\_Tol\textasciicircum{}-1 \\ \hline
\rowcolor[HTML]{FFFFFF} 
Bot\_T\textasciicircum{}1 Bot\_Ph\textasciicircum{}1 & Top\_Tol\textasciicircum{}-1 Bot\_Tol\textasciicircum{}1 \\ \hline
\rowcolor[HTML]{FFFFFF} 
Top\_Tol\textasciicircum{}-1 Bot\_Tol\textasciicircum{}1 & sin(Bot\_MCH) \\ \hline
sin(Top\_Tol) &  \\ \hline
sin(Bot\_MCH) &  \\ \hline
\begin{tabular}[c]{@{}l@{}}Bot\_Tol\textasciicircum{}1\\   P22\textasciicircum{}-1\end{tabular} &  \\ \hline
\begin{tabular}[c]{@{}l@{}}Bot\_MCH\textasciicircum{}-1\\   Phenol\textasciicircum{}1\end{tabular} &  \\ \hline
\rowcolor[HTML]{FFFFFF} 
\begin{tabular}[c]{@{}l@{}}Bot\_T\textasciicircum{}-1\\   Bot\_Ph\textasciicircum{}-1\end{tabular} &  \\ \hline
\rowcolor[HTML]{FFFFFF} 
\begin{tabular}[c]{@{}l@{}}Top\_Tol\textasciicircum{}1\\   Bot\_MCH\textasciicircum{}1\end{tabular} &  \\ \hline
\rowcolor[HTML]{FFFFFF} 
\begin{tabular}[c]{@{}l@{}}Top\_Ph\textasciicircum{}1\\   Top\_Tol\textasciicircum{}-1\end{tabular} &  \\ \hline
\rowcolor[HTML]{FFFFFF} 
sin(Bot\_T) &  \\ \hline
sin(P22) &  \\ \hline
\end{longtable}

\end{document}